\documentclass[%
 reprint,
 amsmath,amssymb,
 aps,
prstab,
]{revtex4-1}

\usepackage{graphicx}
\usepackage{epstopdf}
\usepackage{dcolumn}
\usepackage{bm}
\usepackage{subfigure}

\begin{document}

\preprint{APS/123-QED}

\title{A Simplified Model for Fast Optimization of Free-Electron Laser Oscillator  }

\author{Kai Li$^{1,2}$}
\author{Minghao Song$^{1,2}$}
\author{Haixiao Deng $^{1,}$}%
 \email{denghaixiao@sinap.ac.cn}
 \affiliation{%
 $^1$Shanghai Institute of Applied Physics, Chinese Academy of Sciences, Shanghai 201800, China\\
 $^2$Graduate University of Chinese Academy of Sciences, Beijing 100049, China.
}%

%



\date{\today}

\begin{abstract}
 A simplified theoretical model for free-electron laser oscillator (FELO) simulation which reserves the main physics is proposed. In stead of using traditional macro particles sampling method, the theoretical model takes advantages of low gain theory to calculate the optical power single-pass gain in the undulator analytically, and some reasonable approximations are made to simplify the calculation of power growth in the cavity. The theoretical analysis of single-pass gain, power growth, time-dependent laser profile evolution and cavity desynchronism are accomplished more efficiently. We present the results of infrared wavelength FELO and X-ray FELO with the new model. The results is checked by simulation with GENESIS and OPC which demonstrates the validity of the theoretical model.
\begin{description}
\item[PACS numbers]
41.60.Cr
\end{description}
\end{abstract}

\pacs{Valid PACS appear here}
\maketitle

\section{Introduction}

Free-electron laser (FEL) is a new light source which uses relativistic electron beam passes through the undulator and interaction with the radiation field to generate high brilliant laser pulses. Due to its many advantages, such as rapid and continuous tenability over a wide spectral range, a great deal of interest is attached to it currently. Single-pass high-gain FEL, especially self-amplified spontaneous emission (SASE) schematic, is able to produce high brilliant laser pulses in X-ray region. With the great success of Linear Coherent Light Source\cite{emma2010first} and Spring-8 Angstrom Compact Free Electron Laser\cite{ishikawa2012compact}, several hard X-ray SASE FEL have been built or under construction around the world. Although SASE FEL provides fully transverse coherent and short temporal duration X-ray pulses, it starts from shot noise and produces poorly temporal coherence light\cite{kim1986three}. Numerous schemes, including external seeded FEL\cite{yu1991generation,xiang2009echo,feng2014phase} and self-seeding\cite{amann2012demonstration,geloni2011novel}, have been proposed to solve this problem and improve its performance.

Another promising operating mode is FELO, which works in the low-gain region as well as employs electron beam to pass the undulator multiple times and convert energy to radiation. There are many long wavelength FELO have been established, e.g., the free-electron laser for infrared experiments\cite{oepts1995free} and Duke storage-ring based FEL\cite{yan2016storage}. Also several infrared and THz FELO are under construction, e.g., the first infrared free-electron laser user facility in China\cite{li2016design}. Recently, a promising schematic X-ray free-electron oscillator (XFELO)\cite{kim2008proposal} has been reconsidered thanks to the development of high-reflectivity high-resolution X-ray crystal\cite{shvyd2010high}. XFELO can generate fully temporal coherent and stable peak power laser pulses with the peak brilliance comparable to SASE and the average brilliance several orders of magnitude higher than SASE. However, there are still lots of challenges of FELO scheme, including high repetition electron injector\cite{ostroumov2008development}, heat loading of the Bragg reflection crystal mirror\cite{song2016numerical} and X-ray optics.

 In addition, unlike SASE FEL in which electron beam passes undulator only once, FELO contains an oscillator in which electron beam and optical pulse go through undulator hundreds of times before saturation. Thus tracking the electrons motion and electric field evolution requires lots of calculation and the theoretical analysis and design of FELO become another problem. Although there are some conventional FEL simulation codes such as GENESIS\cite{reiche1999genesis}, GINGER\cite{fawley2001ginger}, by combine with optical codes OPC\cite{van2009time} they can be used to simulate FELO process, these approaches are usually relatively slow and time-consuming. For example, according to\cite{kim2008proposal,dai2012proposal} a complete tracking from the initial spontaneous emission to final saturation of XFELO took about one month. Thus there is a strong scientific demand of a simpler and faster theoretical model, which is able to obtain some primary results, basic performances of XFELO and optimum parameters values with acceptable accuracy.

Traditional simulation approaches mentioned above use macro particles sampling method, which contains tens of thousands of macro particles in each slice, and tracking all of them is laborious and time consuming. In this paper, we propose a new theoretical model which takes advantages of electron distribution function to solve the single-pass gain as a function of electronic field intensity, and some assumptions are used to simplify the calculation of radiation power in order to save time. The new approach reduces the calculation time for a fully tracking of FELO from days to minutes by analyzing single-pass gain, power growth, time-dependent laser profile evolution and cavity desynchronism in a more efficient way. The passage is arranged as following: the second section introduces the three main parts of theoretical model. Then we show two examples: 1.6 $\mu \mathrm{m}$ infrared wavelength FELO and 1 $\mathrm{\AA}$ X-ray FELO. Finally a brief conclusion of this paper is given.

\section{Theoretical model of FEL oscillator}
A FELO facility typically contains multiple mirrors with reflectivity $R$ to form an optical cavity which captures the radiation emitted by relativistic electrons traveling through undulator. In this paper we focus on the two-mirror FEL oscillator and other multi-mirror cases are similar to it. The initial optical field comes from the spontaneous radiation as the electron beam passing through the undulator acts as a seed for the following amplifying. With the carefully synchronism between electron beam and optical pulse, the radiation starting from the shot noise overlaps with electrons and is amplified on successive passes. In fact, the electron beam energy is modulated and converted to radiation pulse in the undulator. In order to ensure the increase of optical power, single-pass gain $G$ should overcome net loss, i.e.,$(1+G)R>1$. The radiation field evolutes in the cavity and for the $(n+1)$th pass at the entrance of undulator
\begin{equation}
E_{n+1}(t)=\left[ E_n(t)g(t)+\delta E(t)\right] R_{total}
\label{eq:eight}
\end{equation}
where $\delta E$ is the spontaneous radiation, $g(t)$ is the gain of optical field and $R_{total}$ is the equivalent reflectivity of two mirrors. The laser field experiences an exponential growth before the gain begin to drop off due to too large energy spread, and the intensity approaches to saturation and remains unchanged finally. After hundreds of passes through the undulator, the laser pulse saturates. And the output power keeps steady while the gain in undulator equals to round-trip total loss in the cavity. The main process laser pulse undergoes during one round-trip can be divided into two main parts: the interaction with electron beam in the undulator and the reflections of mirror at the two sides of cavity. In order to clarify the new FELO model three elementary procedures are analyzed and relative parameters are calculated as following.

\subsection{Gain calculation}
Due to its relatively small number of undulator periods, FEL Low-Gain theory is suitable for analyzing the increase of optical power. According to the Low-Gain theory\cite{boscolo1982gain,boscolo1985classical}, the evolution of electrons distribution and thus the gain of laser pulse are solved analytically. We derive formulas in one dimension approximation, and focus on the longitudinal component of laser field and electrons coordinate. The motion of single electron in the phase space $(\theta , \eta)$ is described by ``pendulum equation''\cite{huang2007review}
\begin{eqnarray}
\frac{d\theta}{dz}=2k_u\eta \label{eq:one1}\\
\frac{d\eta}{dz}=-\frac{\epsilon}{2k_u{L_u}^2}\sin\theta
\label{eq:one}
\end{eqnarray}
where we introduce the field strength parameter
{\setlength\abovedisplayskip{0.3cm}
\setlength\belowdisplayskip{0.4cm}
\begin{equation*}
\epsilon=\frac{eE_0K[JJ]}{{\gamma_r}^2mc^2}k_u{L_u}^2
\end{equation*}}
The electrons distribution function $\rho(z;\eta,\theta)$ at point $z$ is governed by the continuity equation
\begin{eqnarray}
\frac{\partial \rho}{\partial z}+\dot{\theta} \frac{\partial \rho}{\partial \theta}+
\dot{\eta} \frac{\partial \rho}{\partial \eta}=0
\label{eq:two}
\end{eqnarray}
where $\dot{x}=\frac{dx}{dz}$, substituting Eq.~(\ref{eq:one1})(\ref{eq:one}) into Eq.~(\ref{eq:two}) and using scaled parameters
\begin{equation}
\begin{split}
z'=\frac{\sqrt{\epsilon}}{L_u}z\\
\eta'=\frac{2k_uL_u}{\sqrt{\epsilon}}\eta
\end{split}
\label{eq:three}
\end{equation}
yield the following partial derivation equation
\begin{equation}
\frac{\partial \rho}{\partial z'}+\eta ' \frac{\partial \rho}{\partial \theta}+
 \sin\theta \frac{\partial \rho}{\partial \eta'}=0
\label{eq:four}
\end{equation}
Assuming the initial distribution of electron beam fulfils Gaussian function with scaled energy spread  $\sigma_{\eta'}$ and scaled energy deviation $\eta'_0$, the solution can be found by the method of characteristics at the end of undulator is
\begin{widetext}
\begin{equation}
\rho(z';\eta',\theta)=\frac{1}{2\pi}\frac{1}{\sqrt{2\pi}\sigma_{\eta'}} \times \mathrm{exp}\left\{ -\frac{1}{2\sigma_{\eta'}^2}\left[ \frac{\eta' \mathrm{cn}(z';C)-\sin\theta \, \mathrm{sn}(z';C)\mathrm{dn}(z';C)}{1-\cos^2
\frac{\theta}{2}\,\mathrm{sn}^2(z';C)}-\eta'_0 \right]^2 \right\}
\label{eq:five}
\end{equation}
\end{widetext}
where $C^2=\frac{{\eta'}^2}{4}+\cos ^2\frac{\theta}{2}$. The interaction between electrons and laser pulse meets the law of conservation of energy, and thus the power gain of laser is
\begin{equation}
G=\sqrt{m_ec^2K[JJ]{k_u}^{-1}}\frac{I}{c\beta}\frac{1}{2\pi \Sigma^2}\frac{1}{\varepsilon_0 {E_0}^{3/2}}\langle \ \Delta \eta \rangle
\label{eq:six}
\end{equation}
where $2\pi \Sigma^2$ is the cross section of electron beam, $\varepsilon_0$ is the dielectric constant of vacuum, and $\langle \Delta \eta \rangle$ is the average change of $\eta$ in one slice which can be calculated by density function integration. Note that the influence of electron beam emittance can be involved by replacing the energy deviation with equivalent relative energy spread
\begin{equation}
\frac{\sigma'_E}{E_0}=\sqrt{\left( \frac{\sigma_E}{E_0} \right)^2 +\left( \frac{\varepsilon \lambda_u}{4\lambda \beta} \right)^2}
\label{eq:seven7}
\end{equation}

\subsection{Cavity model}
\begin{figure}
\includegraphics[width=8cm]{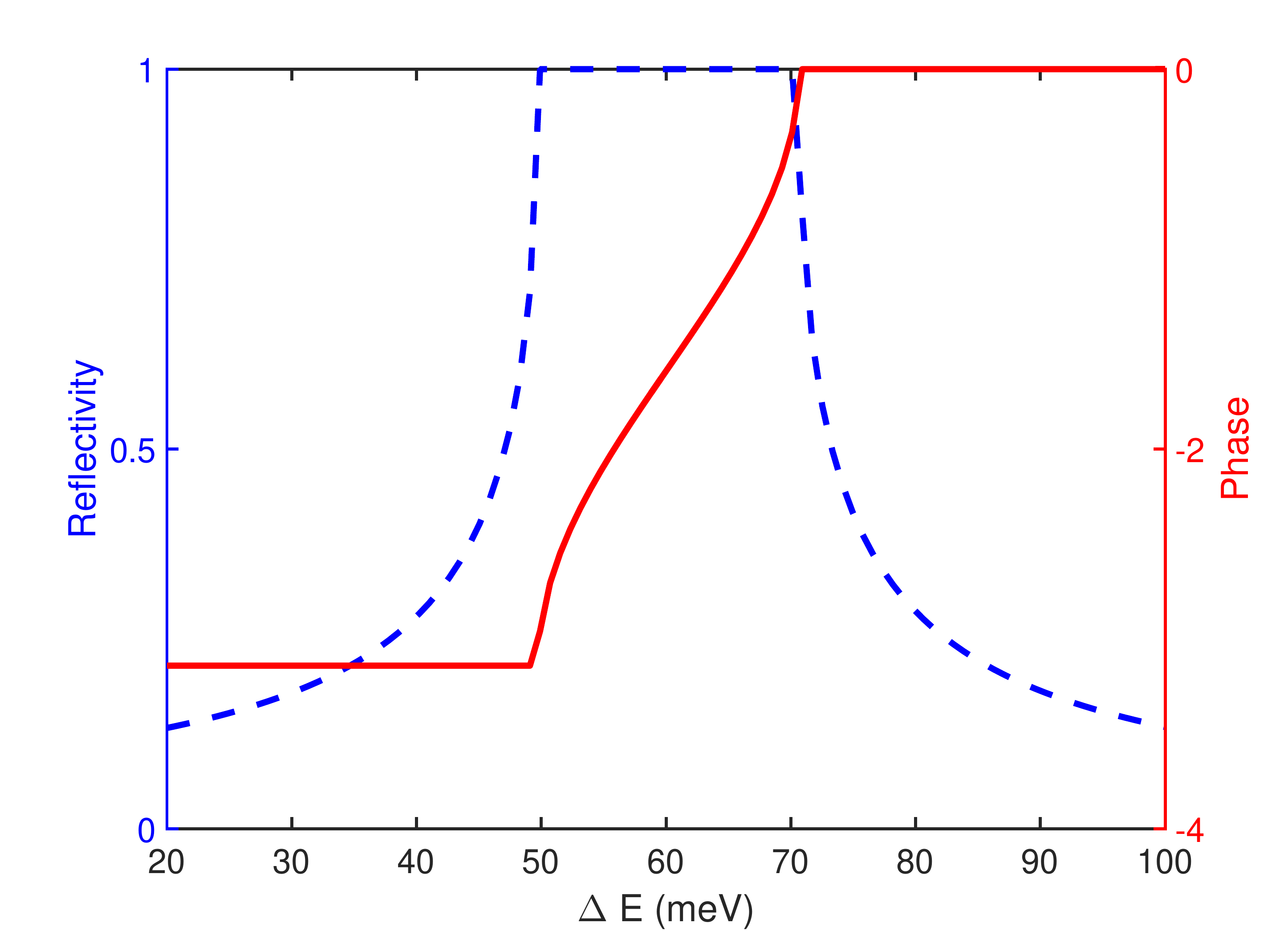}
\caption{\label{fig:RR}The complex reflectivity of Bragg crystal at various incident photon energy deviation from Bragg energy.}
\end{figure}

The FELO facility typically contains a mirror at each side of the FELO cavity, which forms an oscillator to trap the optical pulse. For infrared wavelength light, metal mirrors are utilized due to its broadband reflectivity and high thermal conductivity. The interaction between radiation and mirror can be approximated by a normal reflection without deformation of optical pulse. However, it is more complicated for XFELO which exploits Bragg crystal reflection. The high-reflectivity bandwidth for crystal mirror is relatively much narrow so that the optical pulses are cut off in the frequency domain and deformed in the temporal space\cite{shvyd2012spatiotemporal,lindberg2012time}. Although the reflectivity of crystal mirror is depend on lots of factors, such as X-ray incident angle, profile of pulse and thickness of crystal, to illustrate the main properties of Bragg backscattering, we assume the crystal to be semi-infinite and non-absorbed, and the symmetry Bragg backward scattering is chosen. In this way, the complex reflectivity is simplified as\cite{lindberg2011performance}
\begin{equation}
r(y)=\left\{
\begin{array}{r@{\quad}l}
y-\sqrt{y^2-1} & \mathrm{if} \: y > 1\\
y-i\sqrt{1-y^2} & \mathrm{if} \: |y| \leqslant 1 \\
y+\sqrt{y^2-1} & \mathrm{if} \: y < -1
\end{array}
\right.
\label{eq:seven}
\end{equation}
where $y=\frac{1}{|\chi_H|}\left[ \frac{2(E-E_H)}{E_H}+\chi_0 \right]$, $E_H$ is the Bragg energy and $\chi_0$ and $\chi_H$ are  Fourier components of the dielectric susceptibility of
the crystal. Fig.~\ref{fig:RR} shows the reflectivity of diamond crystal C(4,4,4) in symmetry
Bragg backscatter at various incident photon energy deviation $\Delta E= E-E_h$. The shift of
curve cental from zero results from the fact that peak reflectivity located at an energy slightly different from the Bragg energy.

\subsection{Initial start-up}
The FELO laser start from the electron beam synchrotron radiation in the undulator, which is chaotic in spectrum and randomly in phase. This initial shot noise optical field acts as a seed which is trapped in the optical cavity and amplified on the successive passes. According to \cite{schmuser2014free} an electron passing through an undulator with $Nu$ periods produces a wave train
\begin{equation}
E_0(t)=\left\{
\begin{array}{l@{\quad}l}
E_0\exp(-i\omega t) & \mathrm{if} \: -T/2<t<T/2,\\
0 & \mathrm{otherwise}.
\end{array}
\right.
\end{equation}
where $T=N_u \lambda/c$ is the time for electron travel through the undulator. And the power generated is
\begin{equation}
P_{rad}=\frac{e^2c\gamma^2K^2k_u^2}{12\pi\varepsilon_0}.
\end{equation}
For an electron pulse contains $N$ electrons which is shorter than one wave train (i.e., including only one wave packet), the total electric field is
\begin{equation}
E(t)=E_0\exp(-i\omega t)\sum\limits_j^N \exp(i\phi_j).
\end{equation}
where $\phi_j$ is the initial phase of electric field which is related to the relative location of electron in the bunch. The time-average field power $U\varpropto E_0^2\left| b \right|^2$ with $b=\sum_j^N \exp(i\phi_j)$. The dimensionless variable $\xi=b^2/\left< b^2 \right>$ obeys the simple exponential probability distribution
\begin{equation}
P(\xi)=\exp(-\xi).
\label{eq:ini_pow}
\end{equation}
which means the electric field intensity is proportional to square root of the number of electrons in a coherent length \cite{saldin1998statistical}.

However, in the numerical model, we choose a Gaussian profile electron beam which is far longer than the optical wave train and contains $M$ wave packets. For this long electron bunch, the power probability distribution is more complex, so we approximate the complex possibility distribution with normal distribution and obtain the initial field power distribution by Gaussian sampling with mean value and standard deviation equal 1 and $1/\sqrt{M}$ respectively. The electric field phase fulfils randomly distribution among $[0,2\pi]$.

According to the FEL resonance condition, the optical field propagates ahead of electron beam one resonant wavelength when travel through one undulator period. By the virtue of several hundred times of pass through undulator and continually slippage, the field in different slices ``communicate" with each other and develops into a high brilliant, monochromatic laser pulse. Due to the chaotic phase at the beginning, the initial gain is relatively small\cite{feng2013slippage}, the influence of slippage is roughly considered as phase averaging of different slices within a coherent length. Besides slippage, the field is shaped by Bragg crystal mirror reflection heavily for XFELO and this is expected to improve longitudinal coherence of the laser. Finally, as the laser power increases and single-pass gain drops off, FELO transforms to saturation and maintain a steady state.

\section{Infrared FELO simulation}

\begin{figure}
\includegraphics[width=8cm]{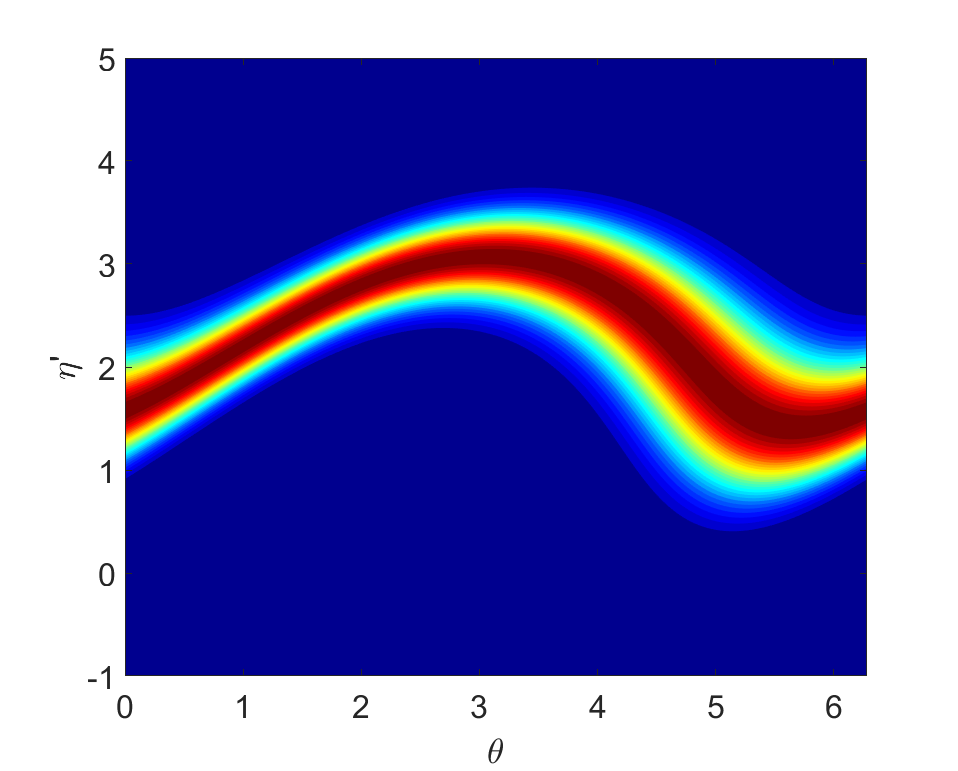}
\caption{\label{fig:contour} The electron density distribution function in phase space of one slice.}
\end{figure}
\begin{figure}
\includegraphics[width=8cm]{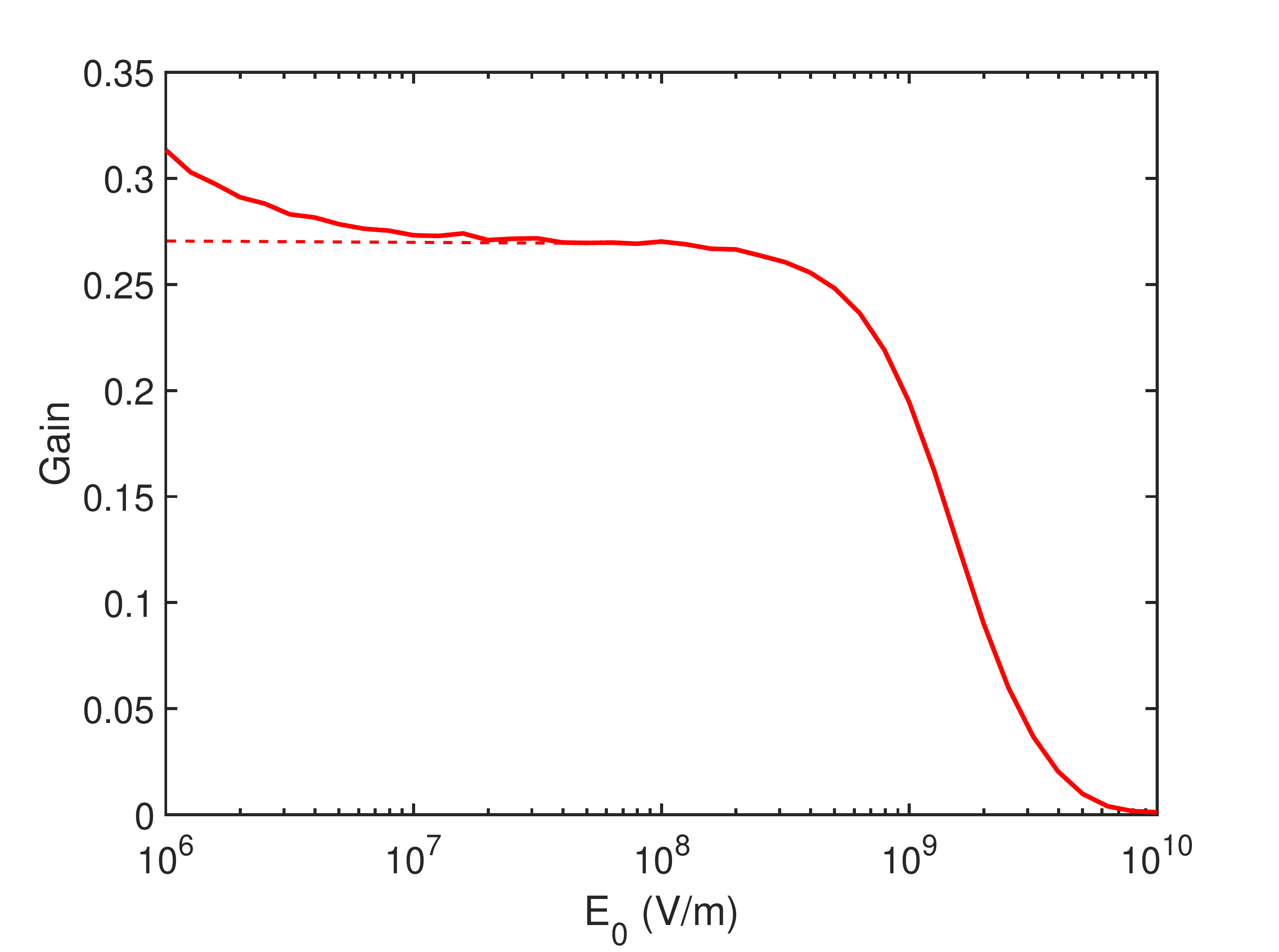}
\caption{\label{fig:gainfun} The gain as a function of field intensity.(The dash line is the expected constant gain function curve in the low-gain theory at small electric field intensity.)}
\end{figure}

\begin{table}
\caption{\label{tab:table1}%
Simulation parameters for infrared FELO.
}
\begin{ruledtabular}
\begin{tabular}{lcr}
\textrm{Parameter}&
\textrm{Value}&
\textrm{Unit}\\
\colrule
Beam energy $E_0$ & 80 & MeV\\
Slice relative energy spread $\sigma_\eta$ & 0.2\% & \\
Normalized emittance $\varepsilon_n$ & 10 & $\mu$m-rad\\
Peak current $I$ & 200 & A\\
Electron beam charge $Q$ & 100 & pC \\
Electron bunch length (FWHM) $\sigma_e$ & 0.5 & ps\\
Undulator period $\lambda_u$ & 45 & mm\\
Number of undulator $N_u$ & 16 & \\
Laser wavelength $\lambda$ & 1.6 & $\mu \mathrm{m}$\\
Rayleigh length $Z_R$ & 0.35 & m\\
Cavity loss & 1\% & \\
Output coupling efficiency & 6\% & \\
\end{tabular}
\end{ruledtabular}
\end{table}
 \begin{figure}
\includegraphics[width=8cm]{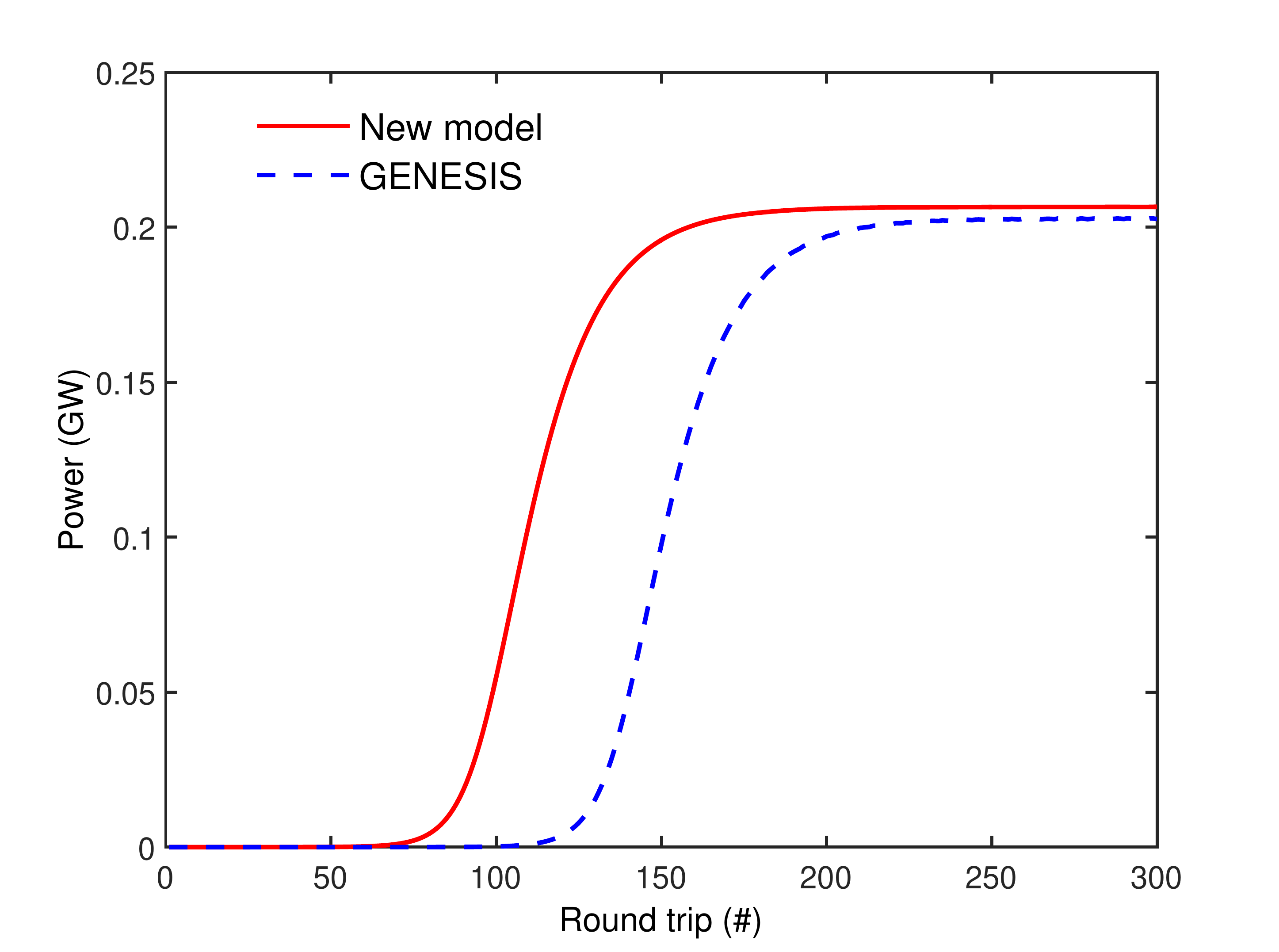}
\caption{\label{fig:N} The enhancement of output laser peak power with various passes $N_{pass}$.}
\end{figure}
\begin{figure*}[t]
  \centering
  \subfigure{\includegraphics[width=4cm]{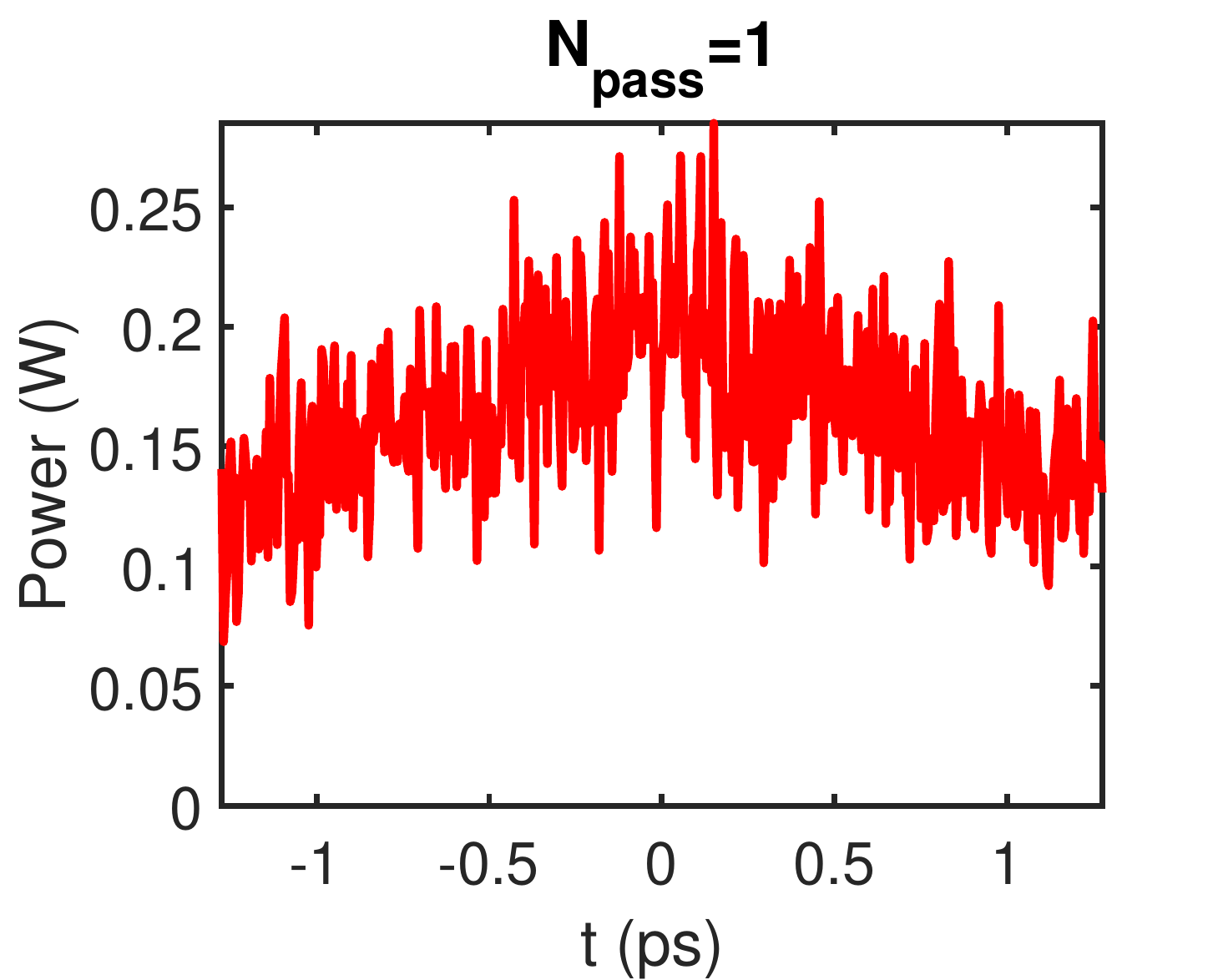}}
  \subfigure{\includegraphics[width=4cm]{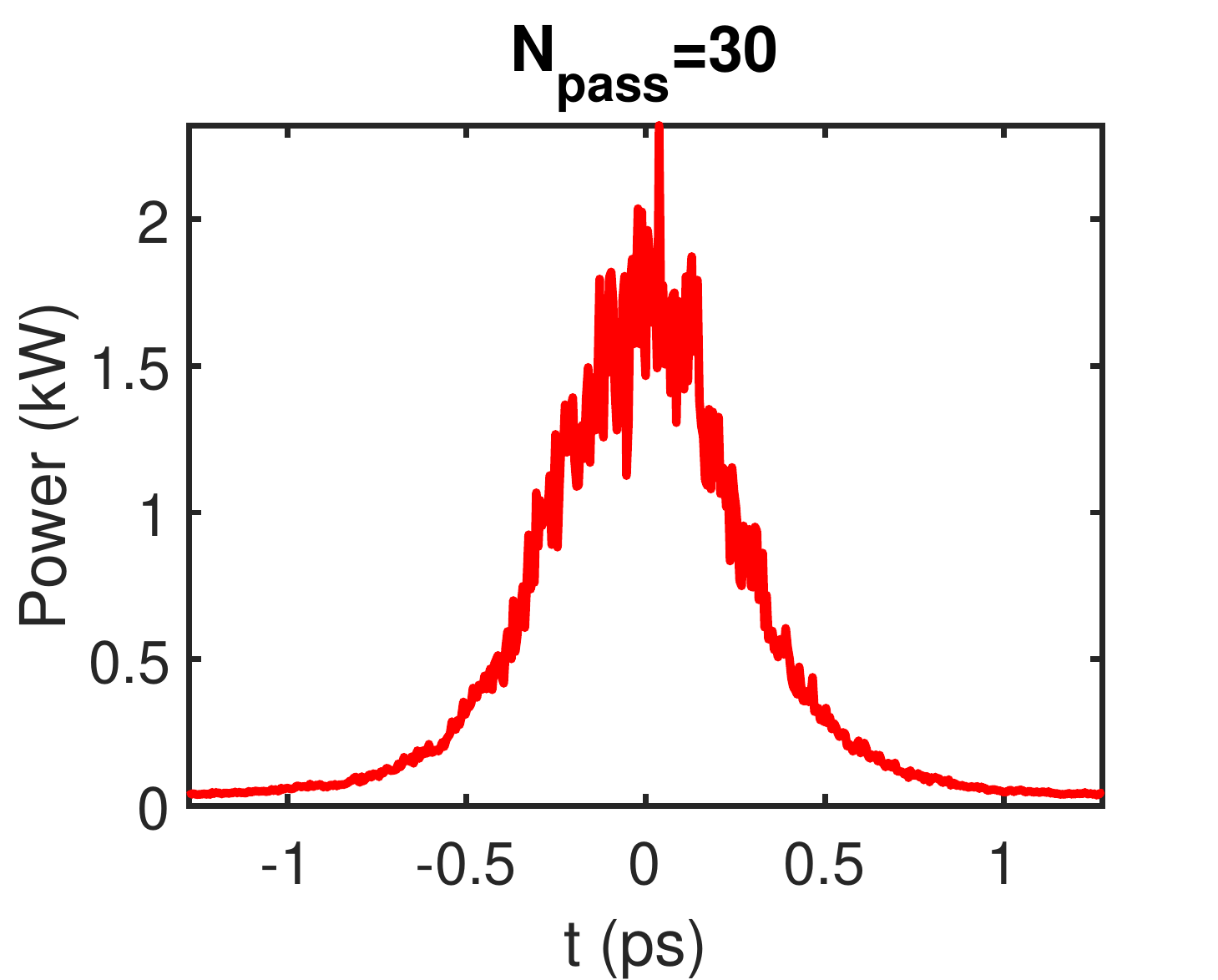}}
  \subfigure{\includegraphics[width=4cm]{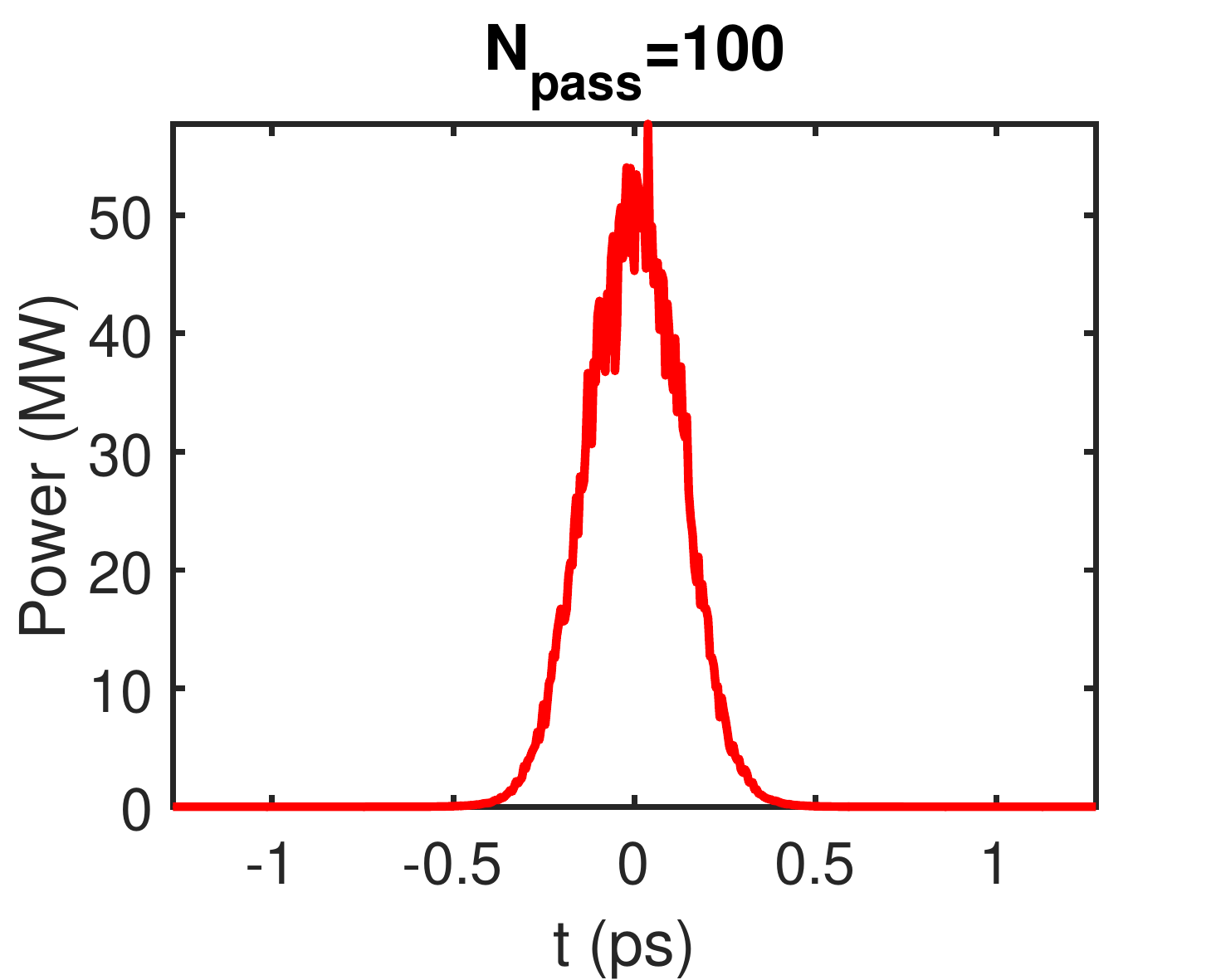}}
  \subfigure{\includegraphics[width=4cm]{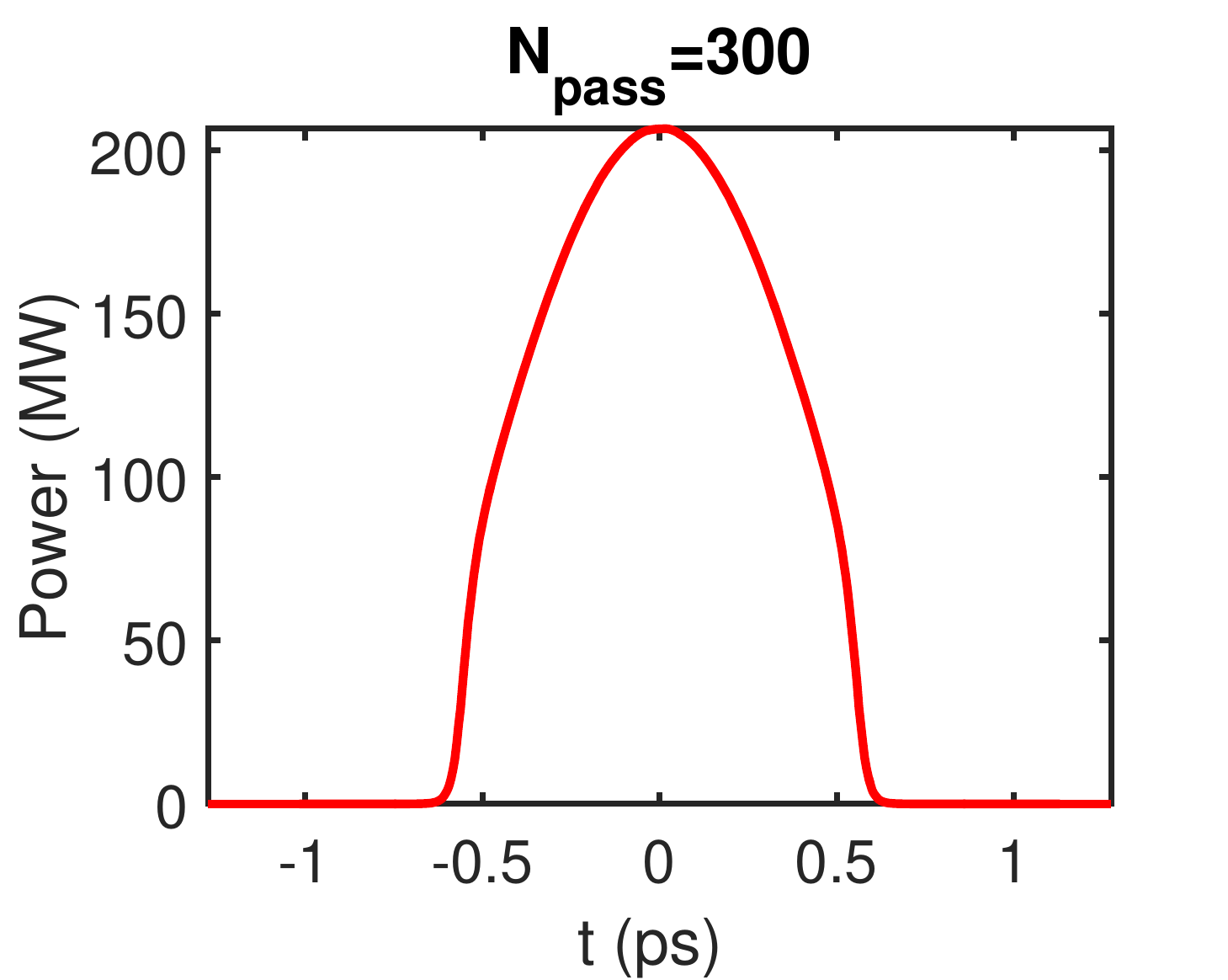}}
  \subfigure{\includegraphics[width=4cm]{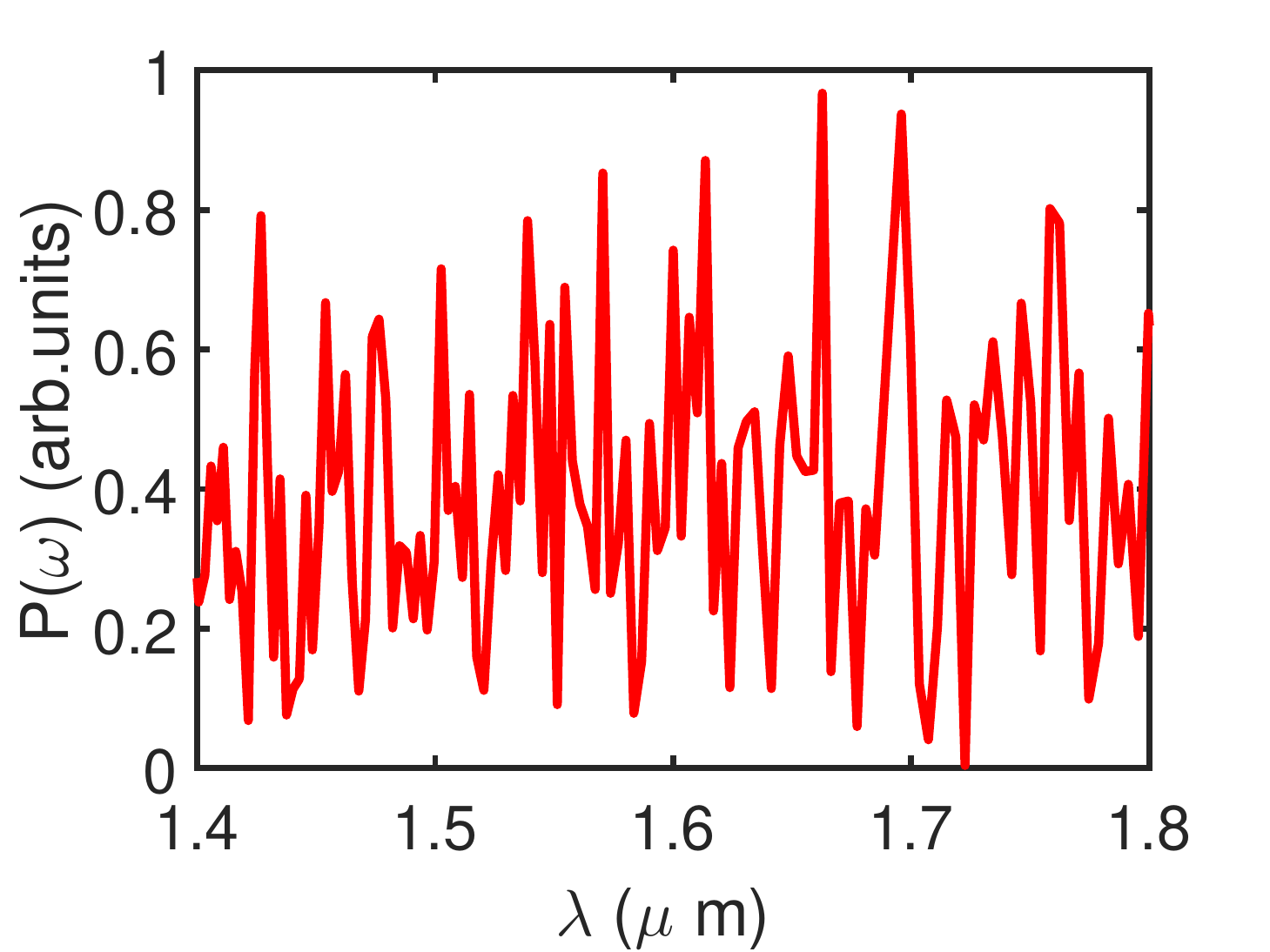}}
  \subfigure{\includegraphics[width=4cm]{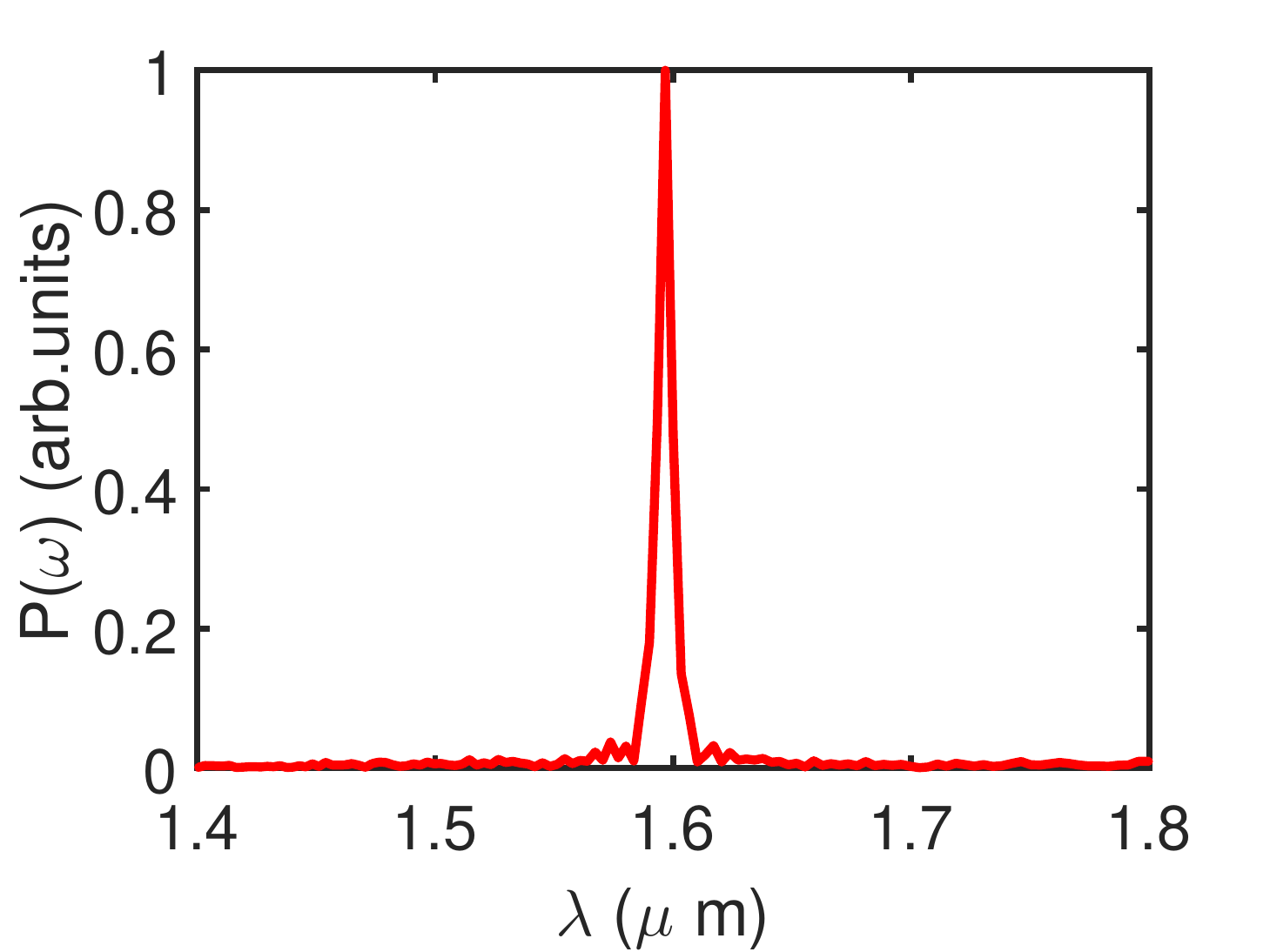}}
  \subfigure{\includegraphics[width=4cm]{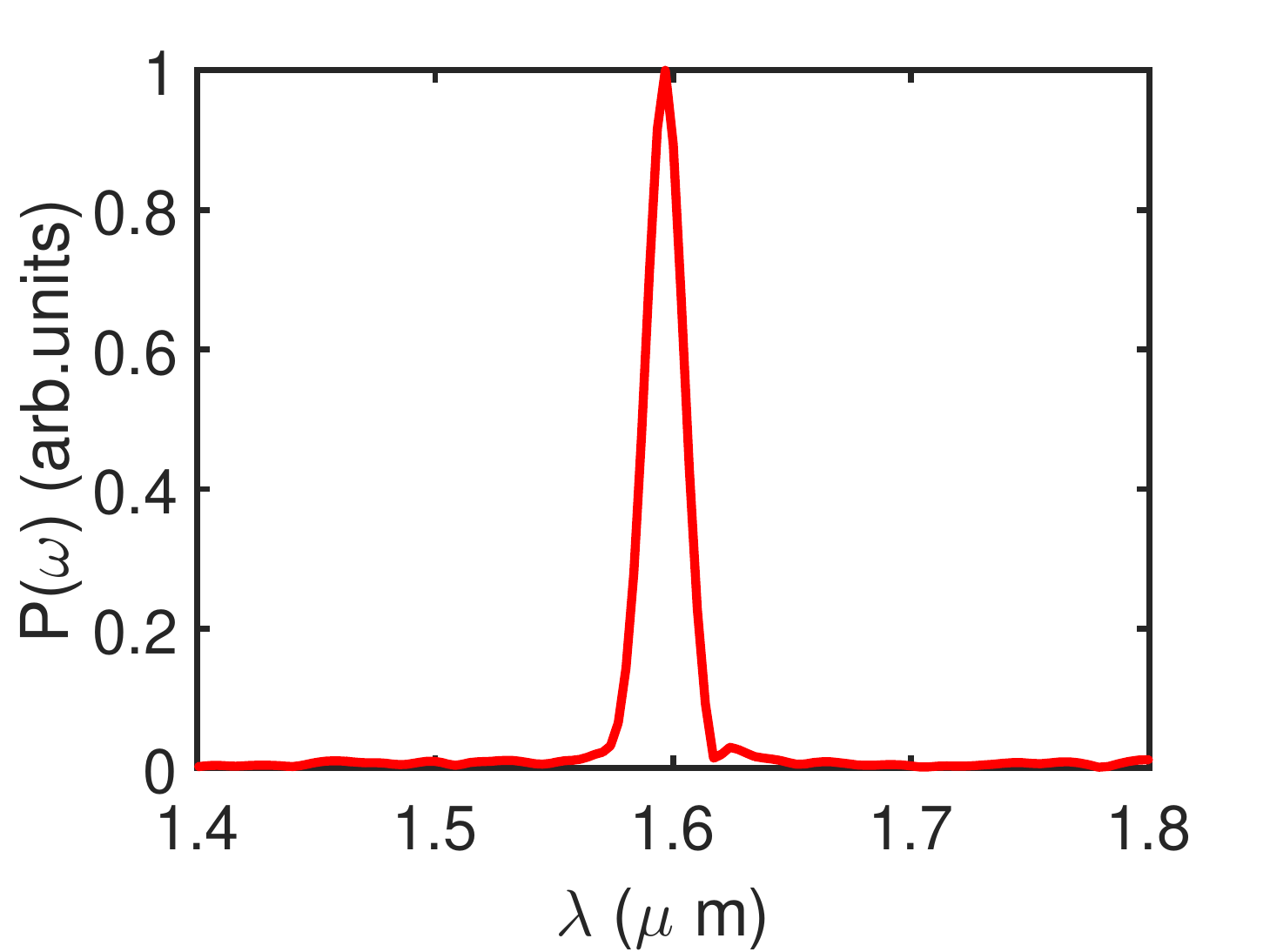}}
  \subfigure{\includegraphics[width=4cm]{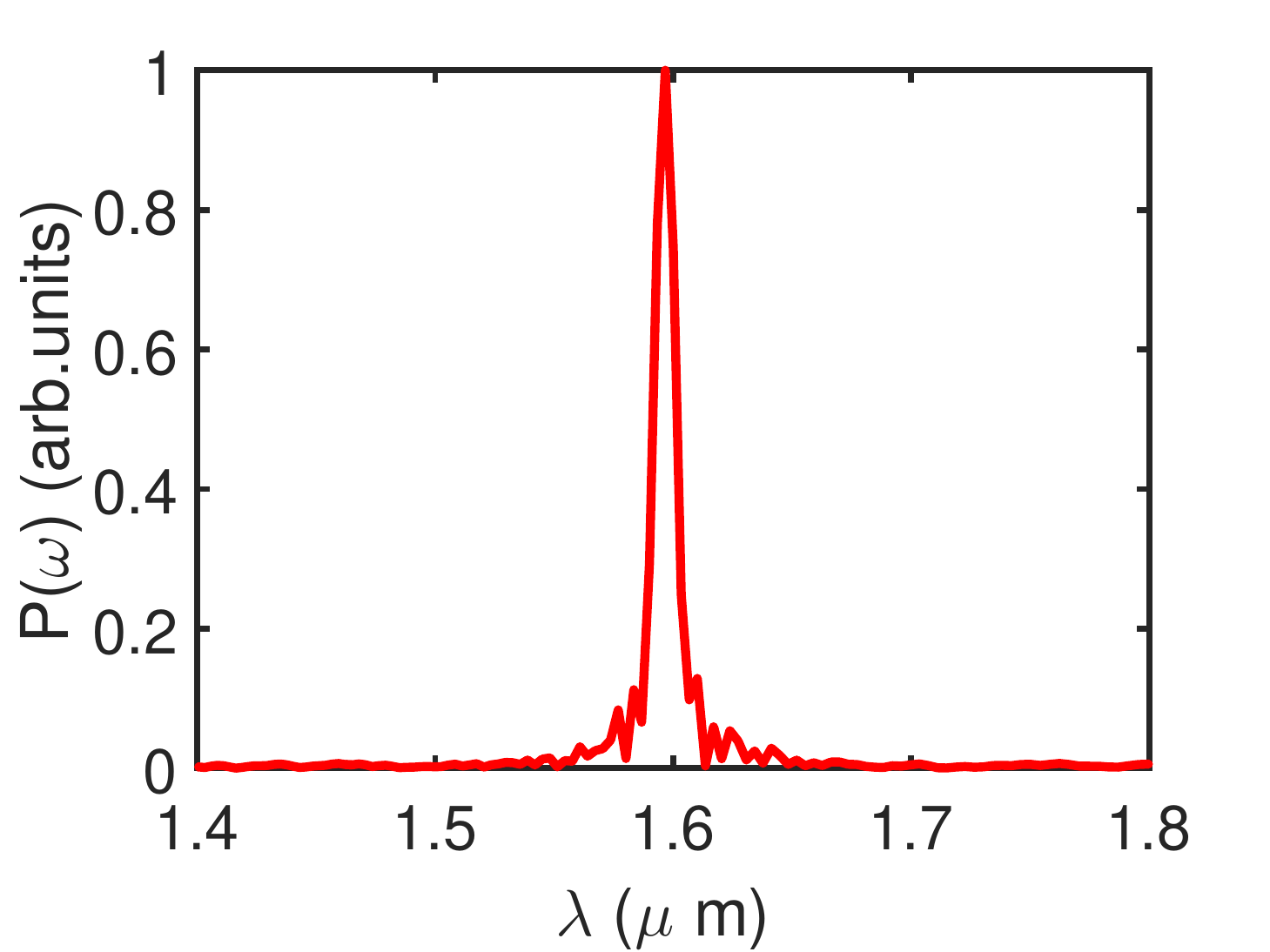}}
 \caption{\label{fig:sandt} Snapshots of output radiation pulse for a typical infrared FELO at
 1.6$\mu m$. The top and the bottom row show the longitudinal pulse temporal profile and corresponding spectrum respectively.}
\end{figure*}
 A typical infrared FELO is investigated using parameters shown in Table.~\ref{tab:table1} which are also presented in\cite{deng2016proposals}. The modulation of electron distribution function in the phase space and the gain degradation as the laser field increase are obtained by Eq.~(\ref{eq:five}). When calculating the gain we assume that the FELO is operated at its optimum electron energy deviation which produces the maximum single-pass gain. And the electron bunch cross section is considered the same as optical beam waist which is determined by Rayleigh length $Z_R$. Fig.~\ref{fig:contour} shows the contour map of electrons distribution in the phase space of one slice. As expected, the initial Gaussian profile is twisted and the electrons rotates in the ``bucket'', which means the electrons energy are modulated and transformed to laser pulse. As a consequence of energy conservation law, the growth of laser power is equal to the loss energy of electron bunch which can be obtained by difference between initial and final total energy using Eq.~(\ref{eq:six}). The sing-pass gain as a function of optical field is given by Fig.~\ref{fig:gainfun}, gain decreases as laser power rises. However, it drops off a little at the beginning which is contradict to the predicted constant gain in low-gain theory. Further investigation reveals that is due to the relative large value of $\eta'$ leads to low accuracy of integration of electron density function at the small radiation power. By refine the mesh grids in the integration process, it can be mitigated.

It is worthy to note that, even refining the mesh grids enhances the accuracy of gain calculation, low gain theory is used to get the gain at low optical intensity while solve the gain by the new model near the saturation. Through the combination of these two approaches, single-pass gain for all electric field intensity can be given effectively and correctly.

To evaluate the evolution of laser pulse profile and spectrum bandwidth in the oscillator, we use Eq.~(\ref{eq:ini_pow}) to get the initial shot noise signal, and adjusted it to nearly 1W for typical infrared FELO case in this paper. The growth of laser power is calculated by Eq.~(\ref{eq:eight}) until it approaches saturation and remain constant. The metal mirrors reflectivity are assumed to be 100\%. The output coupling of the downstream mirror is 6\% and extra 1\% cavity loss due to the mirror absorbing or deflection of light is considered. Fig.~\ref{fig:N} displays the growth of laser pulse after various number of pass. The evolution of both spectral and temporal profile of radiation in the cavity from shot noise to saturation are demonstrated at the upper and bottom row of panels respectively in Fig.~\ref{fig:sandt}. These plots indicate significant temporal and spectral fluctuation during the initial amplify period and become smooth as passing number increases. The FELO finally generate a $1\mathrm{ps}$ laser pulse with narrow spectral bandwidth at $1.6\mathrm{\mu m}$. Excepting a little faster purify of spectrum and earlier boost of laser power, which are due to the roughly treatment of slippage and initial shot noise power, the results are agree well with Ref.~\cite{deng2016proposals} both in temporal and spectrum profile as well as peak intensity. Thus it proves the accuracy of the new theoretical model. In addition, the laser pulse temporal width first narrows and then broadens, which is consist with the analysis in Ref.~\cite{kim1991spectral}.

\begin{figure}
\includegraphics[width=8cm]{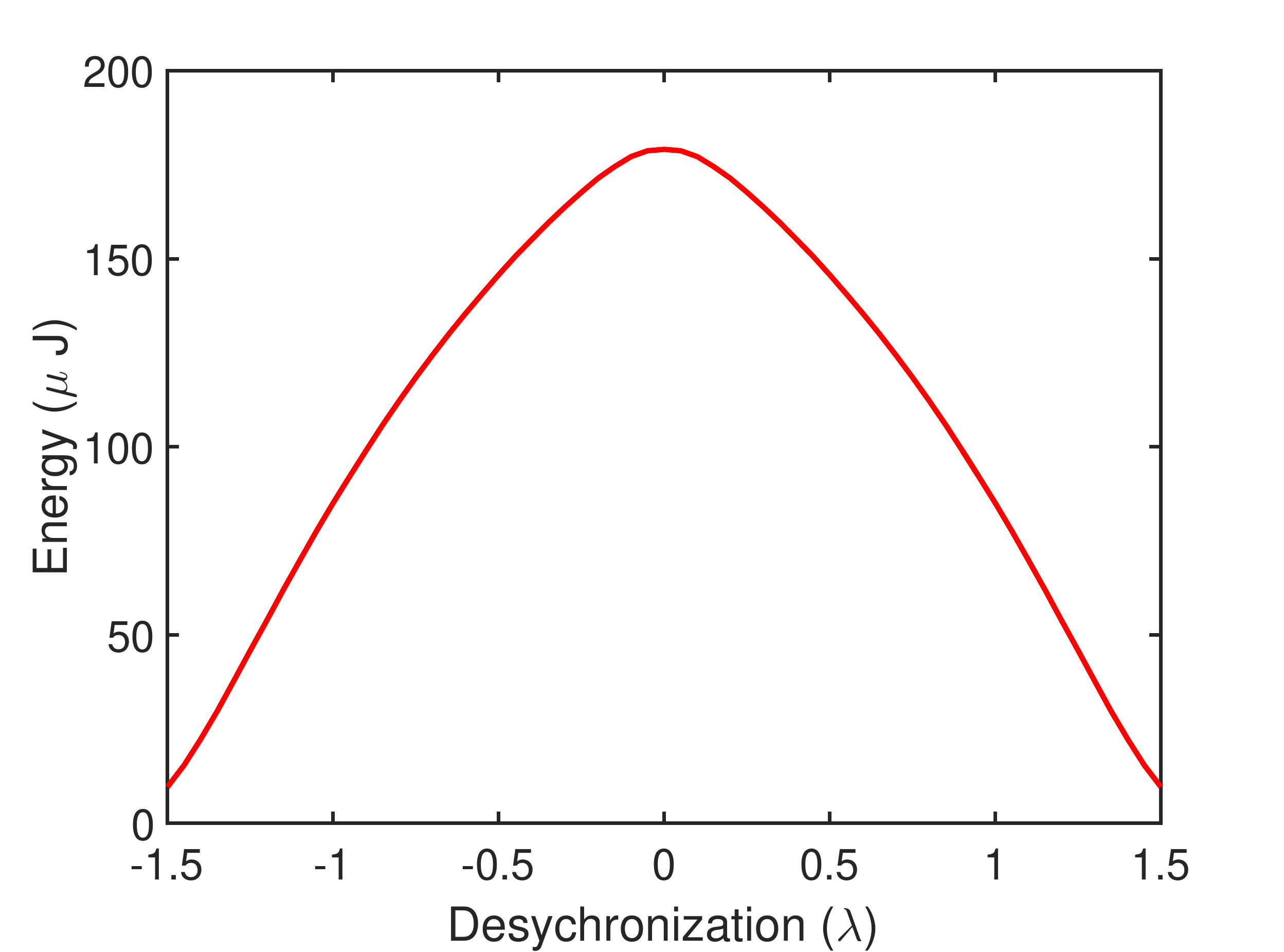}
\caption{\label{fig:shift} The output laser energy as a function of desynchronism.}
\end{figure}

\begin{figure}
\includegraphics[width=8cm]{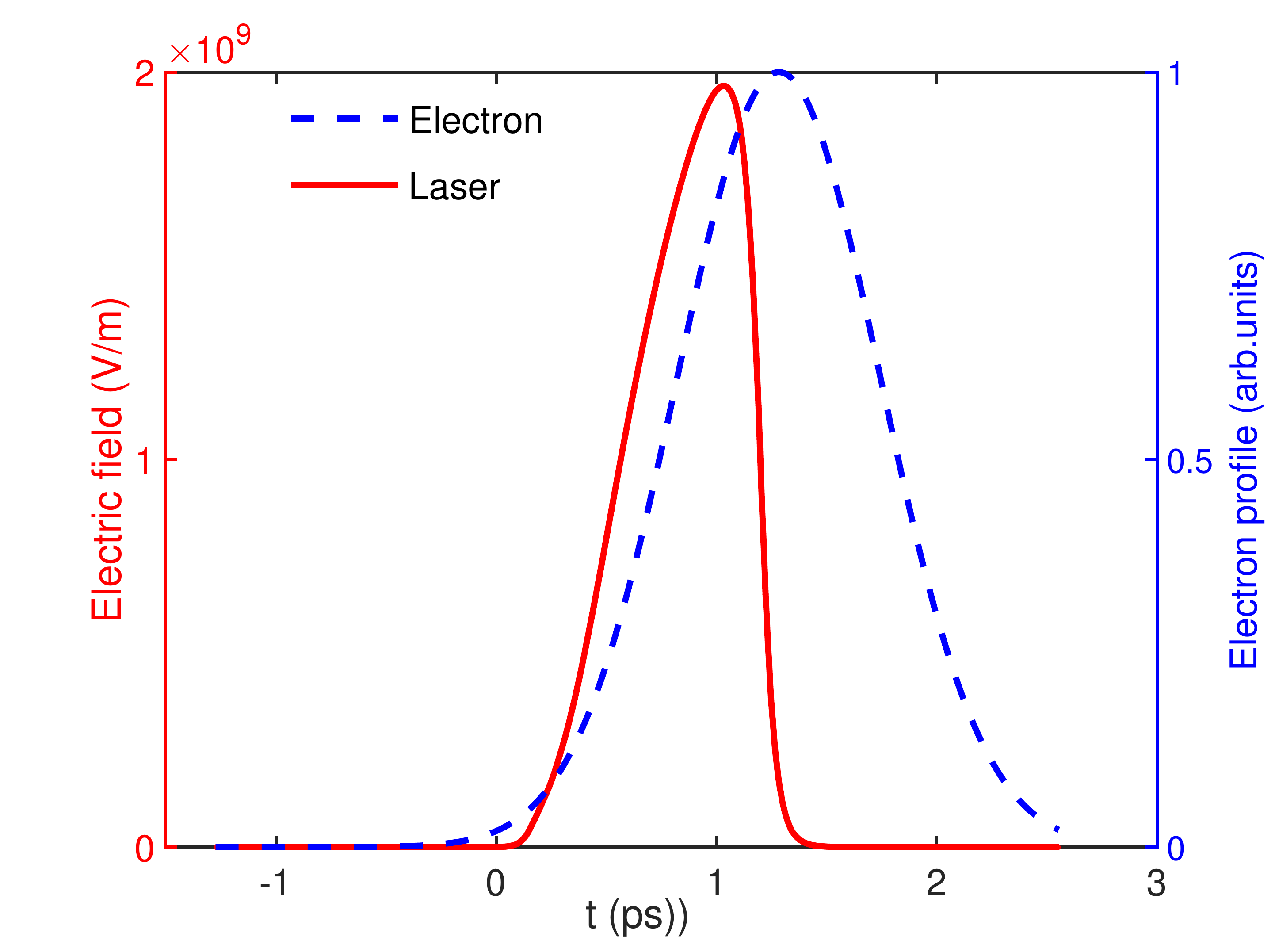}
\caption{\label{fig:shiftprofile} The laser profile at 0.6$\lambda$ cavity desynchronism.}
\end{figure}

It is interesting to note that since the new theoretical model generates results more efficiently, it can be adjusted to investigate the cavity desynchronism in FELO. Due to slippage, the group velocity of optical field is smaller than the speed of light $c$, which lead to the maximum gain appears to lag behind and miss out the largest electron density location. This leads to the degradation of output laser peak power\cite{bakker1994short,knippels1995intense}. It is hard to calculate the group velocity of light precisely in our model, so we ignore the ``lethargy'' effect and assume the electron bunch and laser pulse already at perfectly synchronism. However, we can deliberately shift the electron profile of a different range and investigate its influence to output laser energy. Fig.~\ref{fig:shift} shows the output energy declines when cavity desynchronism increases. Fig.~\ref{fig:shiftprofile} demonstrates a typical light profile at $0.6\lambda$ cavity desynchronism and it is obvious that the pulse tilts towards electron beam slightly.
\section{x-ray FELO simulation}
\begin{table}
\caption{\label{tab:table2}%
Simulation parameters for X-ray FELO.
}
\begin{ruledtabular}
\begin{tabular}{lcr}
\textrm{Parameter}&
\textrm{Value}&
\textrm{Unit}\\
\colrule
Beam energy $E_0$ & 7 & GeV\\
Energy spread $\sigma_E$ & 1.4 & MeV\\
Normalized emittance $\varepsilon_n$ & 0.2 & $\mu$m-rad\\
Peak current $I$ & 10 & A\\
Electron bunch length $\sigma_t$ & 1.0 & ps\\
Undulator period $\lambda_u$ & 17.6 & mm\\
Number of undulator $N_u$ & 3000 & \\
Laser wavelength $\lambda$ & 1.0 & $\mathrm{\AA}$\\
Cavity loss & 5\% & \\
Bragg mirror reflectivity $R$ & 94\% & \\
\end{tabular}
\end{ruledtabular}
\end{table}
\begin{figure}
\includegraphics[width=8cm]{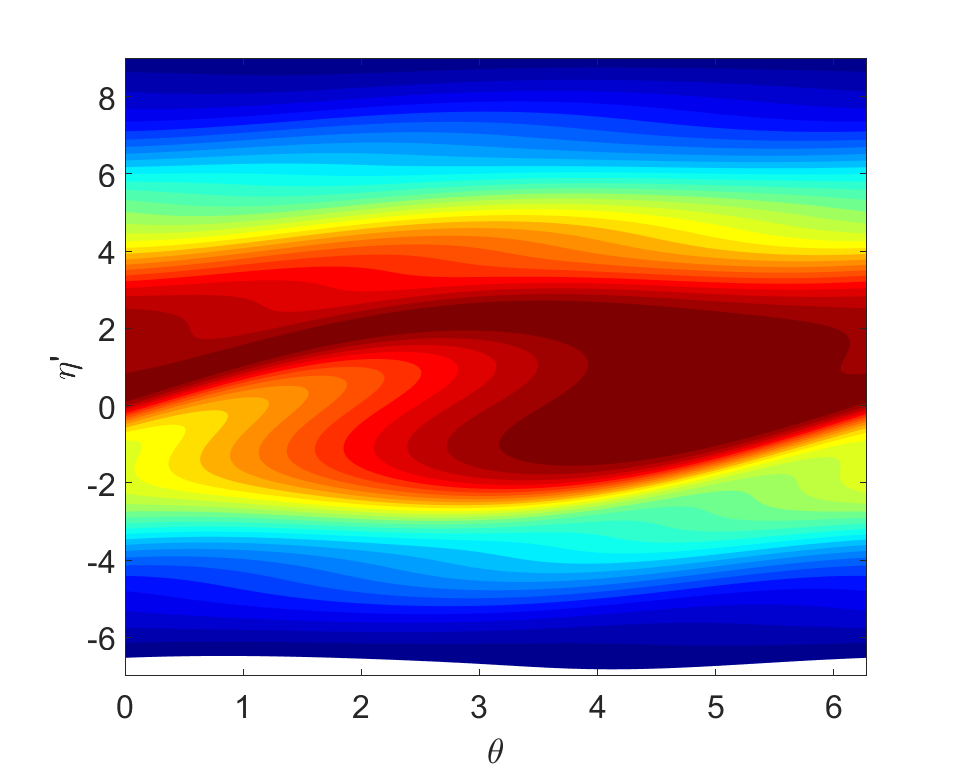}
\caption{\label{fig:xelectrond} The electron density distribution in the phase space of one slice.}
\end{figure}
\begin{figure}
\includegraphics[width=8cm]{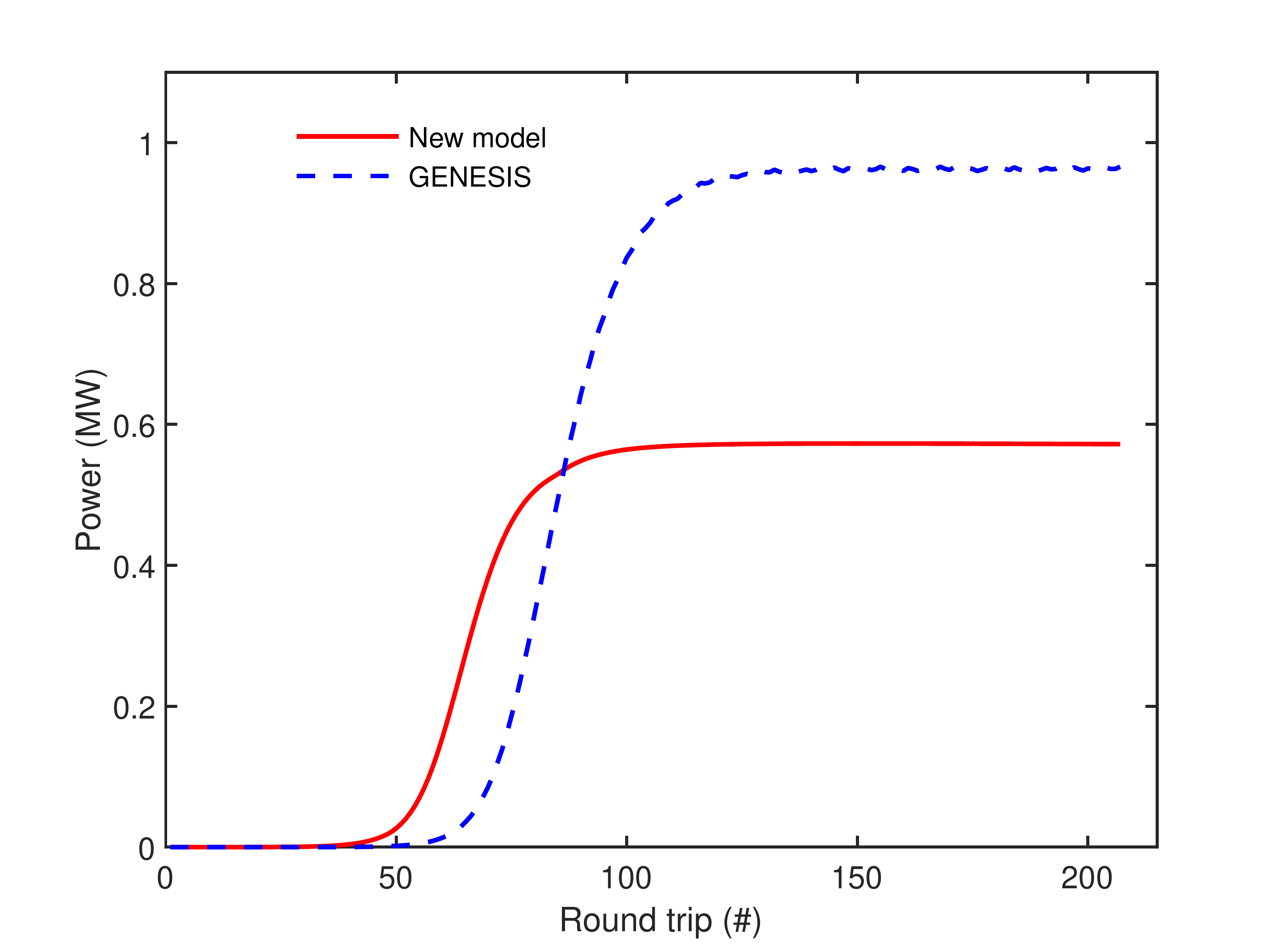}
\caption{\label{fig:xround_trip} The enhancement of output laser peak power with various passes $N_{pass}$.}
\end{figure}
A typical $1\,\mathrm{\AA}$ X-ray XFELO is investigated with the parameters in Table.~\ref{tab:table2}. The optical cavity is built up with diamond crystal mirrors by C(4,4,4) Bragg reflection with the Bragg energy at $12.04\,\mathrm{keV}$ and reflectivity reaches 94\%. The Bragg crystal mirrors filter out the optical frequency beyond its high-reflectivity spectral bandwidth, and in order to maintain enough gain to overcome the round-trip loss and magnify the optical field, the number of undulator periods is chosen to be 3000. In this case the single-pass gain is near 39\% which is given by the theoretical gain calculation method mentioned above. The electron density distribution in the phase space of a slice, when pulse power inside the cavity equals to 0.2MW, is shown in Fig.~\ref{fig:xelectrond}. Due to the relative larger electron energy, the bucket which traps the electron becomes flatter, and the energy modulation is smaller. After a shot periods of struggling as the initial shot noise, the laser power inside the cavity increases as it goes through the undulator again and again and becomes a stable 0.6MW laser pulse output at saturation, which is shown in Fig.~\ref{fig:xround_trip}. The final output power from GENESIS simulation presented with dashed line is nearly twice as large as that from the new model, which is mainly due to the inaccurate evaluation of electron beam and laser cross section as well as the coupling factor between them. Given the approximations used in the analysis process, it is easy to understanding this discrepancy.
\begin{figure*}[t]
  \centering
  \subfigure{\includegraphics[width=4cm]{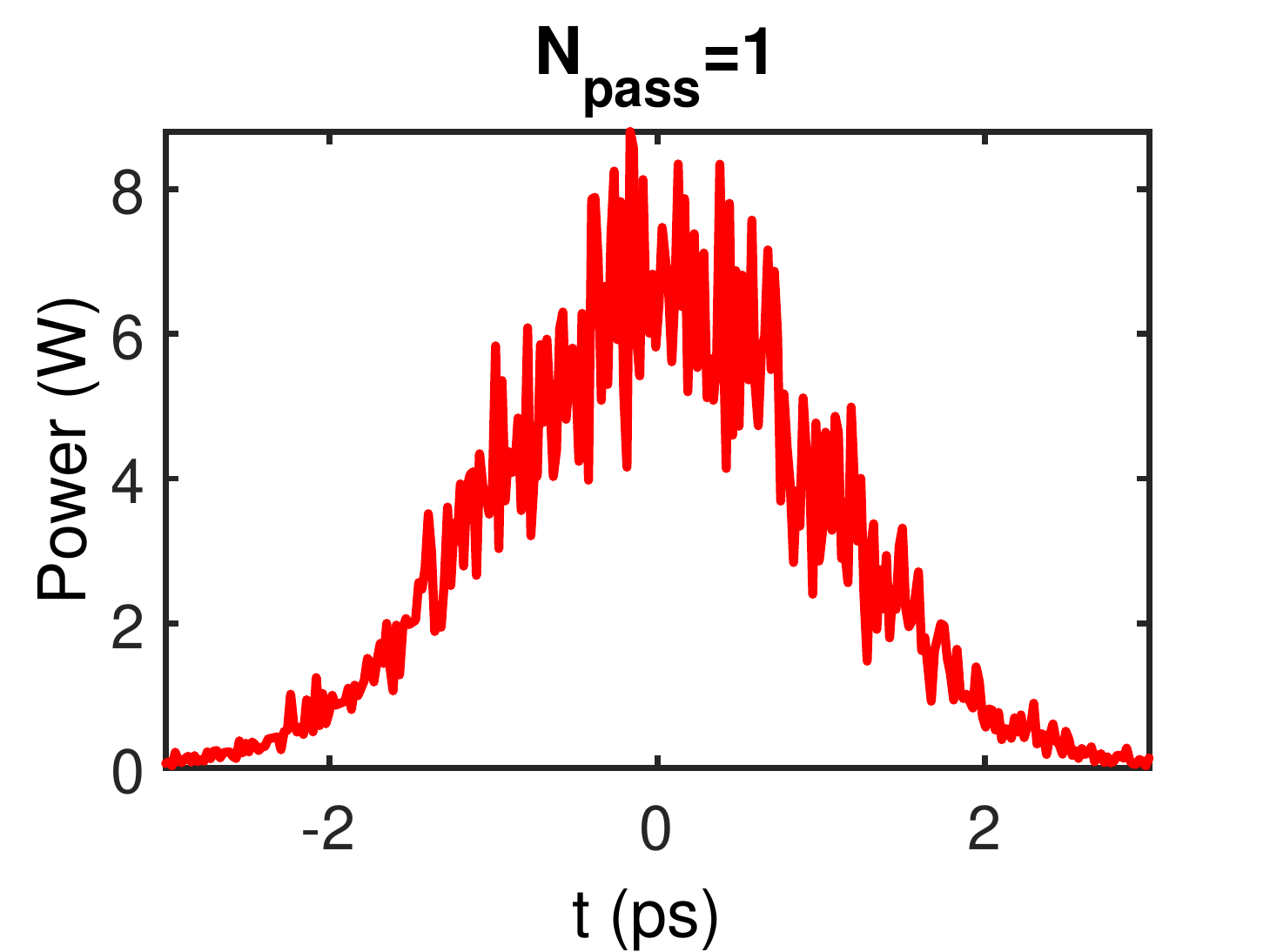}}
  \subfigure{\includegraphics[width=4cm]{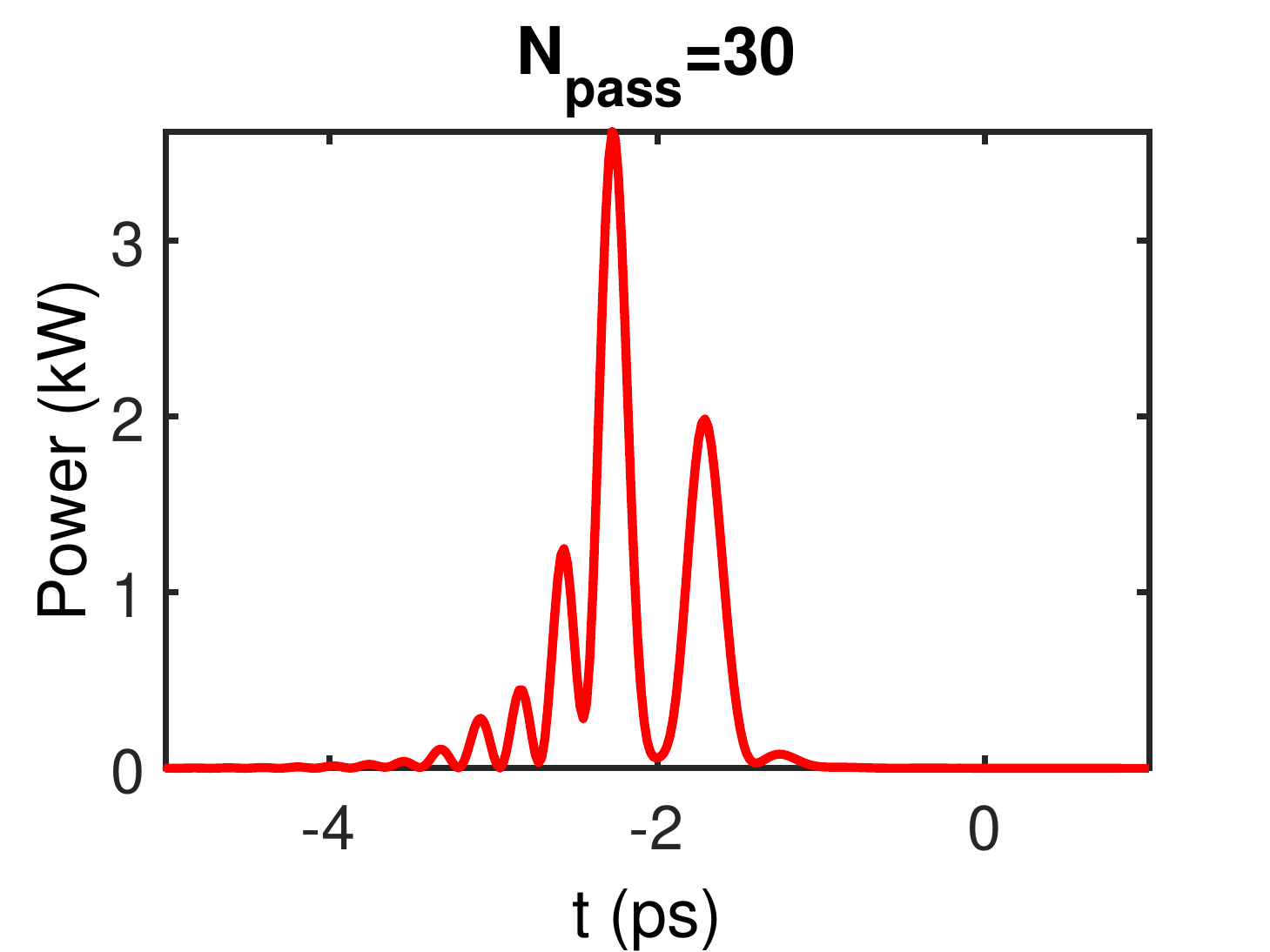}}
  \subfigure{\includegraphics[width=4cm]{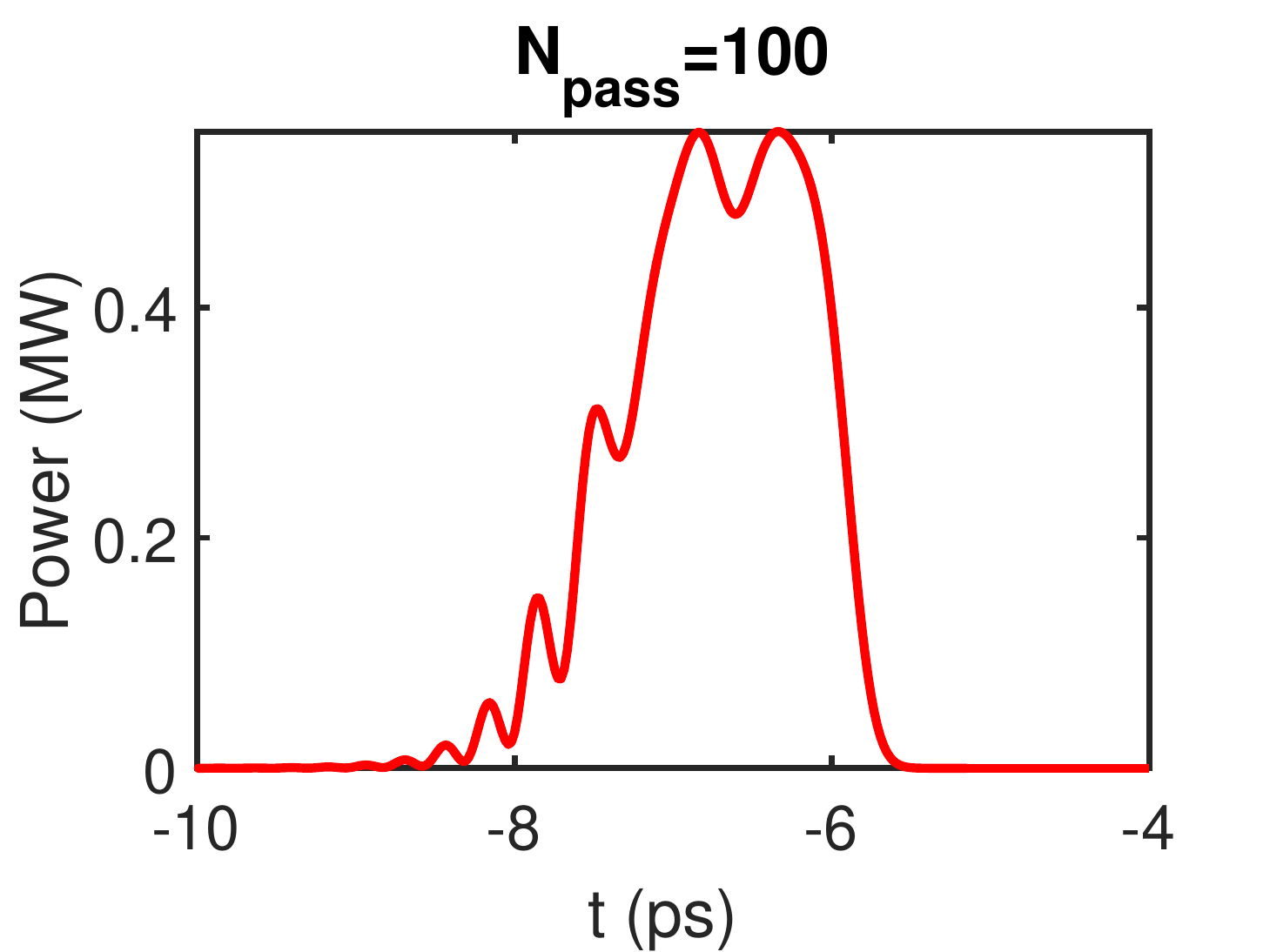}}
  \subfigure{\includegraphics[width=4cm]{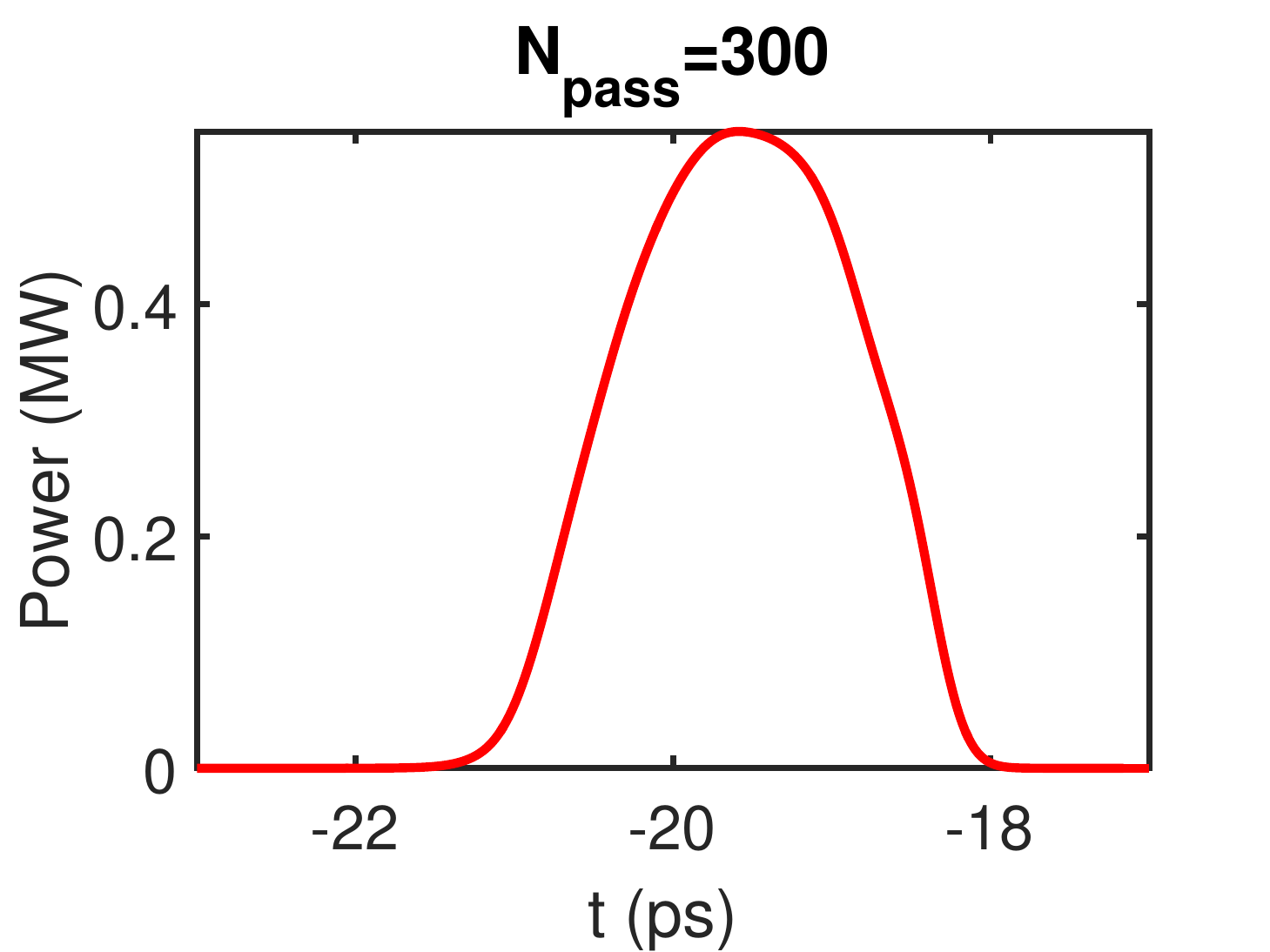}}
  \subfigure{\includegraphics[width=4cm]{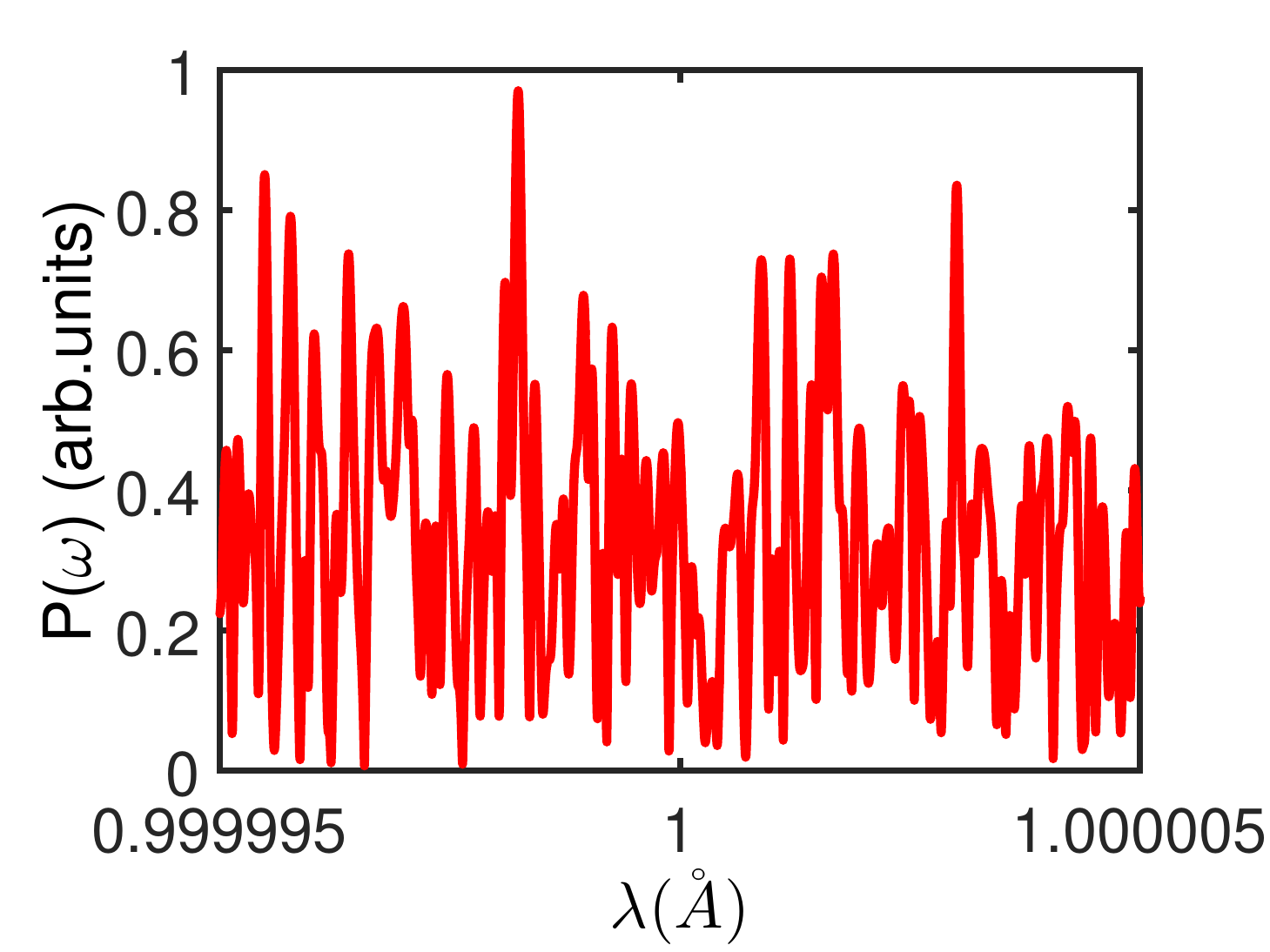}}
  \subfigure{\includegraphics[width=4cm]{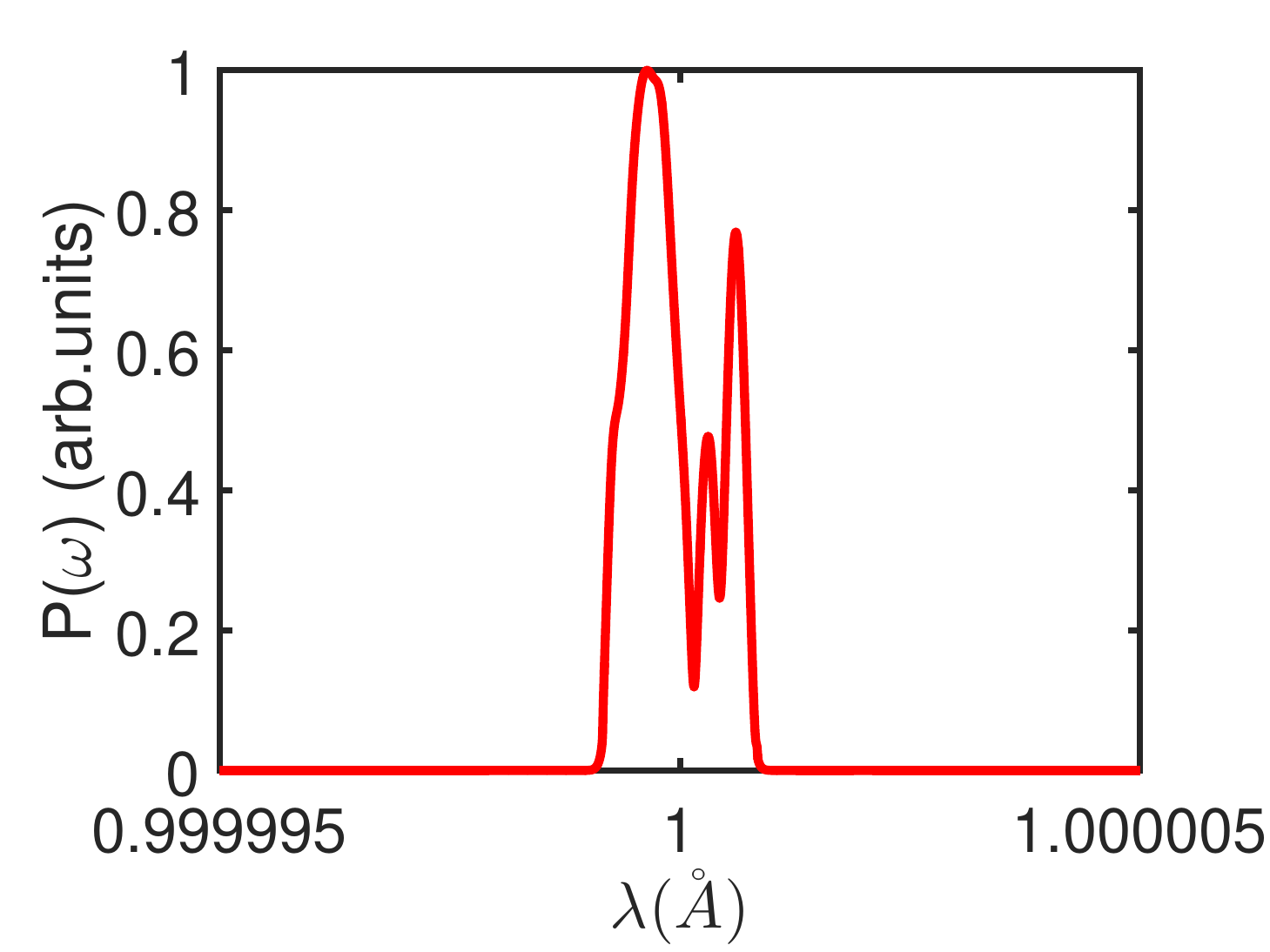}}
  \subfigure{\includegraphics[width=4cm]{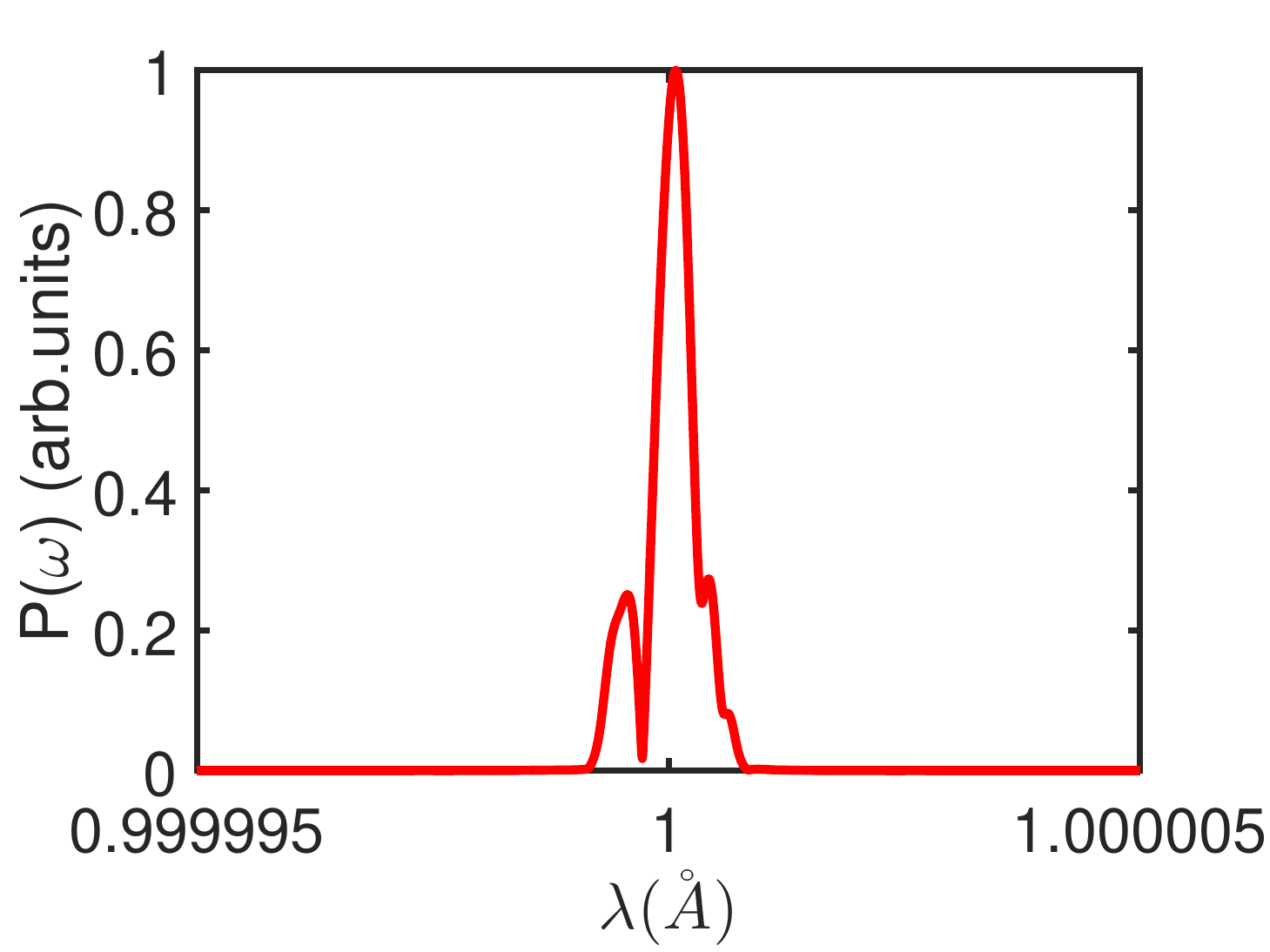}}
  \subfigure{\includegraphics[width=4cm]{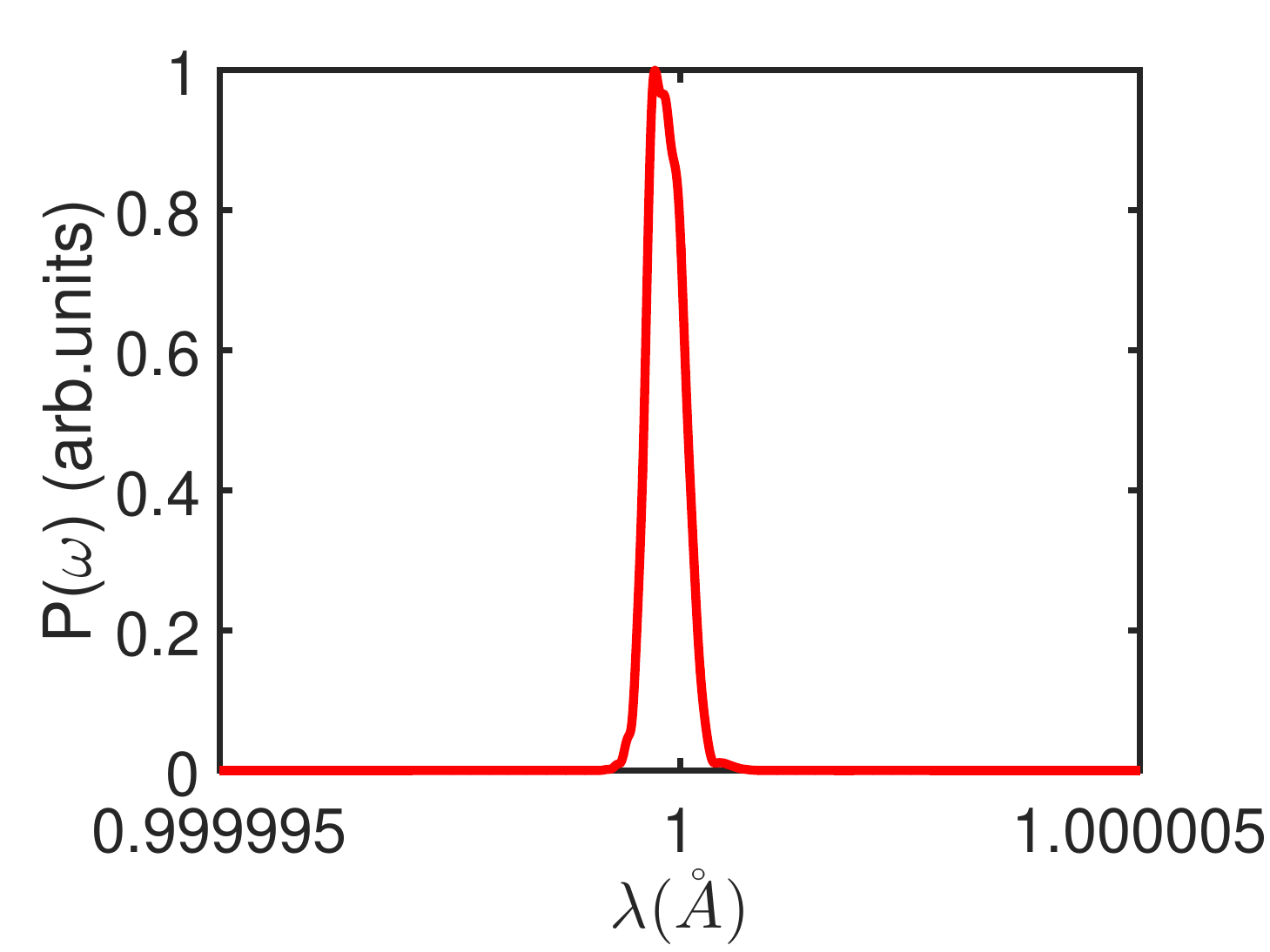}}
 \caption{\label{fig:xsandt} Snapshots of output radiation pulse for a typical X-ray FELO at
 1.0 $\mathrm{\AA}$. The top and the bottom row show the longitudinal pulse temporal profile and corresponding spectrum respectively.}
\end{figure*}
\begin{figure}
\includegraphics[width=8cm]{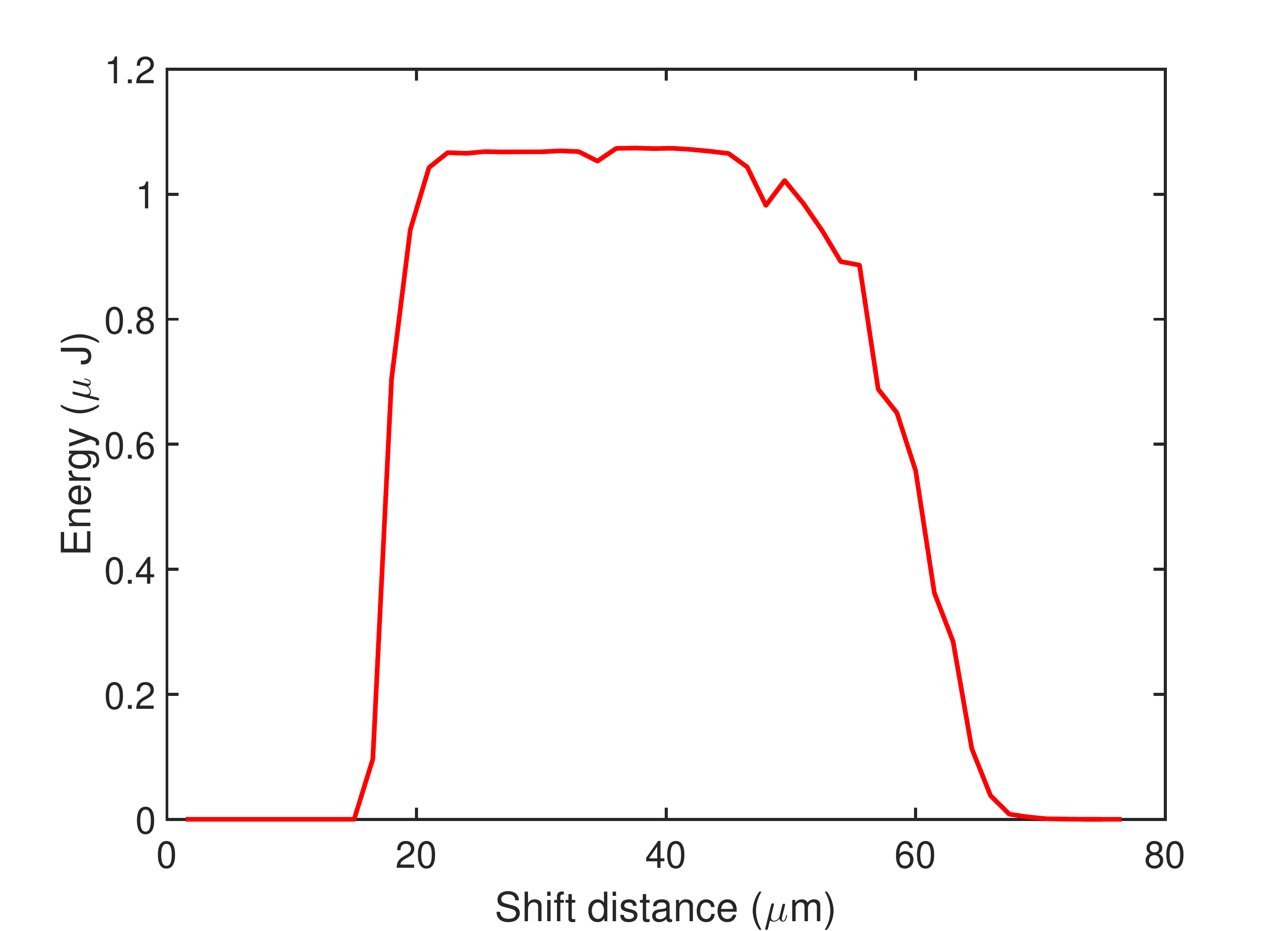}
\caption{\label{fig:xshift} The output laser energy as a function of electron beam shift distance in each pass.}
\end{figure}

The XFELO laser power evolution inside the cavity is solved in the same method as for infrared FELO except that the reflectivity of metal mirrors is replaced by complex reflectivity in Eq.~(\ref{eq:seven}) and the snapshots of output pulse profile and the corresponding spectrum are displayed in Fig.~\ref{fig:xsandt}. The evolution of the laser power and profile are quite similar with the infrared wavelength case. However, due to the spectrum purifying of the crystal mirrors, the output laser spectral bandwidth is much smaller than the previous infrared FELO. And the complex reflectivity of the crystal mirrors causes an extra phase shift of optical field and leads to the pulse slides backward. The ``lethargy'' from the crystal mirrors reflection are much larger than that of the infrared FELO, thus cannot be ignored. In the theoretical model, the electron beam is constantly delayed a distance to overlap with the optical field and the shift distance is equal to 20$\mu$m to maintain the overlap between electron bunch and laser pulse in Fig.~\ref{fig:xsandt}. The output laser FWHM duration is nearly 2ps.

The output total energy as a function of this shift distance is demonstrated in Fig.~\ref{fig:xshift}. The energy increases sharply when the shift distance larger than $18\mu m$ and reduces gradually when it is bigger than $50\mu$m. The fluctuation of laser energy when the electric field struggles to grow up at edge of over-desynchronism is due to the different initial shot noise at each case. Our result shows that the energy remain constant when the shift distance is 30$\mu$m larger than the optimum desynchronism. The flat-top profile of the curve is due the relative narrow high-reflectivity spectral bandwidth of crystal mirrors which leads to the corresponding broad temporal optical profile which acts as a long seed to be amplified. Thus the high intensity output power covers larger desynchronism range in XFELO.

\section{conclusion}
In this paper we proposed a novel theoretical model which is useful for fast optimization of FELO and obtain some results in a shorter period of time. The model solved the partial differential equation theoretically in order to obtain the single-pass gain when FELO is approaching saturation. The oscillator mirrors reflection is considered simply by multiplying the reflectivity. The gain calculation as a separate part takes a few minutes and once the gain function is established as a data base, the oscillator cavity simply calls correspond gain and spends tens of seconds to reaches saturation and returns the needed results.

 We investigated the performance of a 1.6 $\mu$m infrared and a 1 $\mathrm{\AA}$ X-ray FELO by the new approach. The agrement between our results and those from GENESIS simulation proves that the new model is feasible and reliable. Taking the advantages of the higher efficiency of the new model, it is easily adjusted to investigate the cavity desynchronism quickly. The electrons initial distribution function is assumed to be Gaussian function, the following work would be to using the truly electron distribution from accelerator tracking to obtain some practical results and to enhance the accuracy of XFELO theoretical model as well as taking into account the influence of laser pulse heating effects on the Bragg crystal mirrors.

\begin{acknowledgments}
The authors are grateful to Xiaofan Wang, Han Zhang and Mengying Deng for helpful discussions and useful comments. This work was partially supported the National Natural Science Foundation of China (11322550) and Ten Thousand Talent Program.
\end{acknowledgments}

\nocite{*}

\bibliographystyle{apsrev4-1}
\bibliography{biblifile}

\end{document}